\documentclass[conference,twocolumn]{support/IEEEtran}

\usepackage{amsmath}
\usepackage{amsthm}
\usepackage{amssymb}
\usepackage{graphicx}
\usepackage{subfig}
\usepackage{rotating}
\usepackage{flushend}
\usepackage{verbatim}
\allowdisplaybreaks
\usepackage{mathtools}

\usepackage{xcolor}

\newcommand{\argmin}{\mathrm{argmin}}

\newcommand{\logt}{\log_{2}}

\newcommand{\abs}[1]{\left| #1 \right| }

\newcommand{\set}[1]{\left\{ #1 \right\}}

\newcommand{\defn}{: =}

\newcommand{\msf}[1]{\mathsf{#1}}
\newcommand{\mbf}[1]{\mathbf{#1}}
\newcommand{\mcf}[1]{\mathcal{#1}}
\newcommand{\idc}[1]{\msf{1}_{#1} }

\DeclareMathOperator{\markov}{\setlength{\unitlength}{.5cm} \begin{picture}(1,1)  \put(0,.22){\line(1,0){1}}  \put(.5,.22){\circle{.3}}   \end{picture}}

\newtheorem{theorem}{Theorem}
\newtheorem{define}[theorem]{Definition}
\newtheorem{lemma}[theorem]{Lemma}
\newtheorem{cor}[theorem]{Corollary}

\IEEEoverridecommandlockouts

\usepackage{etoolbox}

\providetoggle{arxiv}
\settoggle{arxiv}{true}

\begin{document}

\title{Inner Bound for the Capacity Region of Noisy Channels with an Authentication Requirement}

\author{
\IEEEauthorblockN{Jake Perazzone}
\IEEEauthorblockA{Lehigh University\\
Bethlehem, PA 18015\\
jbp215@lehigh.edu}
\and
\IEEEauthorblockN{Eric Graves and Paul Yu}
\IEEEauthorblockA{US Army Research Lab\\
Adelphi, MD 20783\\ 
ericsgraves@gmail.com, paul.l.yu.civ@mail.mil}
\and
\IEEEauthorblockN{Rick Blum}
\IEEEauthorblockA{Lehigh University\\
	Bethlehem, PA 18015\\
	rb0f@lehigh.edu}
\thanks{This material is based upon work partially supported by the U. S. Army Research Laboratory and the U. S. Army Research Office under grant numbers W911NF-17-1-0331 and W911NF-17-2-0013 and by the National Science Foundation under grants ECCS-1744129 and CNS-1702555.}
}

\maketitle

\begin{abstract}
The rate regions of many variations of the standard and wire-tap channels have been thoroughly explored.
Secrecy capacity characterizes the loss of rate required to ensure that the adversary gains no information about the transmissions.
Authentication does not have a standard metric, despite being an important counterpart to secrecy.
While some results have taken an information-theoretic approach to the problem of authentication coding, the full rate region and accompanying trade-offs have yet to be characterized.
In this paper, we provide an inner bound of achievable rates with an average authentication and reliability constraint.
The bound is established by combining and analyzing two existing authentication schemes for both noisy and noiseless channels.
We find that our coding scheme improves upon existing schemes.
\end{abstract}

\section{Introduction}
Authentication, or the ability to verify the identity of the sender of received transmissions, is crucial in secure communications.
It is especially important in the wireless channel where malicious parties have easy access to all nodes and can attempt to intercept messages and impersonate legitimate senders.
While cryptographic authentication methods are very practical, they are limited to computational complexity as the basis for security.
The first information theoretic analysis of authentication was done by Simmons \cite{Auth} for the noiseless channel in which it was shown that an opponent's attack success probability is lower bounded by $ 2^{-n\kappa/2} $ when the legitimate parties share a key of length $ n\kappa $.

Similar to coding for secrecy in the wire-tap channel, an authentication constraint can be added to a channel code. 
In \cite{Maurer00a}, Maurer likened authentication to a binary hypothesis test for whether a received message is authentic versus inauthentic.
Naturally then, an authentication code would have a decoder that groups certain observations as authentic and others as inauthentic in addition to mapping to possible codewords.
A larger authentic set would allow for an increase in rate since fewer observations would be thrown out as inauthentic, but would allow an adversary to more easily send messages that would be falsely authenticated.
It is because of this that the additional constraint on the code should lead to a trade-off between rate and authentication capabilities in our inner bound. 

In \cite{lai2009authentication}, Lai et. al. presented a code for noisy channels with authentication capabilities and concluded that if the main channel is not less noisy than the adversary, it is possible to limit the attack success probability to $ 2^{-n \kappa} $ with a shared key $ K $ of length $ n\kappa $. 
Although it was shown that the communication rate is unaffected if $ n\kappa $ is small, their analysis is only concerned with cases where $n\kappa$ is a constant independent of $n$.
Gungor and Koksal \cite{gungor2016basic} explored a more general problem and presented an inner bound on the achievable rate with error and erasure exponents for impersonation and substitution attacks both with and without a shared key.
We consider the model of~\cite{lai2009authentication}, while not requiring a constant $n\kappa$ and determine an inner bound that improves upon Gungor and Koksal's coding scheme. Of interest is that the coding scheme can be decomposed into two separate coding schemes, one for source authentication and one for channel authentication.
A direct proof is given for the region while the converse is left for future work. If the converse were true, it would prove that authentication under the operational requirements is a limited resource, and that this resource and the message rate must linearly share the channel's capacity.

Our contributions are as follows.
First, for all DM-ASC$(t,q,1)$, a substitution channel, defined in Section \ref{sec:model}, we give an inner bound on the trade-off between the rate $r$, the key rate $\kappa$, and the average type I error exponent $\alpha$, when the average probability of message error, $\epsilon$, must go to zero with block length $n$ going to infinity.
The average type I error exponent is a measure of authentication ability and is defined in Section \ref{sec:metrics}. It should be noted that this measure of authentication subsumes both the ``impostor'' and ``substitution'' attack.
Our inner bound is characterized in terms of (in principal) computable information theoretic measures in the form of an inner bound.
The derived region subsumes the results of Lai et. al. \cite{lai2009authentication} in which only an asymptotically vanishing key rate is considered.
The inner bound is also a strict improvement over the bounds found in Gungor and Koksal \cite{gungor2016basic}.
Our scheme benefits from higher communication rates and less key leakage.

\iftoggle{arxiv}{}{Due to space limitations the proofs can be found in~\cite{perazzone2018inner}.}

\section{Notation, Model, and Metrics}\label{sec:notations}

\subsection{Notation}
Random variables and their realizations will be denoted by uppercase and lowercase letters, respectively.
The support set of a random variable and other sets are denoted by a calligraphic font.
An $ n $-length sequence of random variables, realizations, or sets will be denoted by superscript $ n $.
So, $ X^n $ is a $n$-length sequence of random variables which may take on values $ x^n \in \mathcal{X}^n.$ The probability $X = x$ is denoted $ \Pr (X=x) $, or $ p_{X}(x)$, and even $ p(x )$ when clear.
The probability of a set is written as $ p_{X}(\mcf{A})=\sum_{x \in \mcf{A}} p(x)$, assuming $ \mcf{A} \subseteq \mcf{X} $ where the set will often be omitted from the summation notation when it is clear, i.e., $ \sum_{x} $. 
The set of all probability distributions on a certain set, say $\mcf{X}$, is denoted by $\mcf{P}(\mcf{X})$.
Similarly, the set of probability distributions of $\mcf{Y}$ conditioned on $\mcf{X}$ is denoted as $\mcf{P}(\mcf{Y}|\mcf{X})$.
The set $\mcf{P}(\mcf{Y}\gg \mcf{X})$ represents a special subset of $\mcf{P}(\mcf{Y}|\mcf{X})$, where if $v \in \mcf{P}(\mcf{Y} \gg \mcf{X})$ for any $y \in \mcf{Y}$, there exists at most one $x \in \mcf{X}$ such that $v(y|x) > 0$. Note, for random variables $X,Y,Z$, if $p_{Y|X} \in \mcf{P}(\mcf{Y}\gg \mcf{X})$, then $X,~Y,~Z$ form a Markov chain, $X \markov Y \markov Z$.	
A superscript of $\otimes n$ will denote the $n$-fold product distribution of $v$.

The use of $O$ will refer to the Bachmann-Landau notation.
When there is a range of possible values for $ O $, we will use $\dot =$ to denote it.
Throughout the paper, the order will only be dependent on continuous functions of the cardinalities of the support sets.

\subsection{Model}\label{sec:model}
\begin{figure}
\centering
\includegraphics[scale=.35]{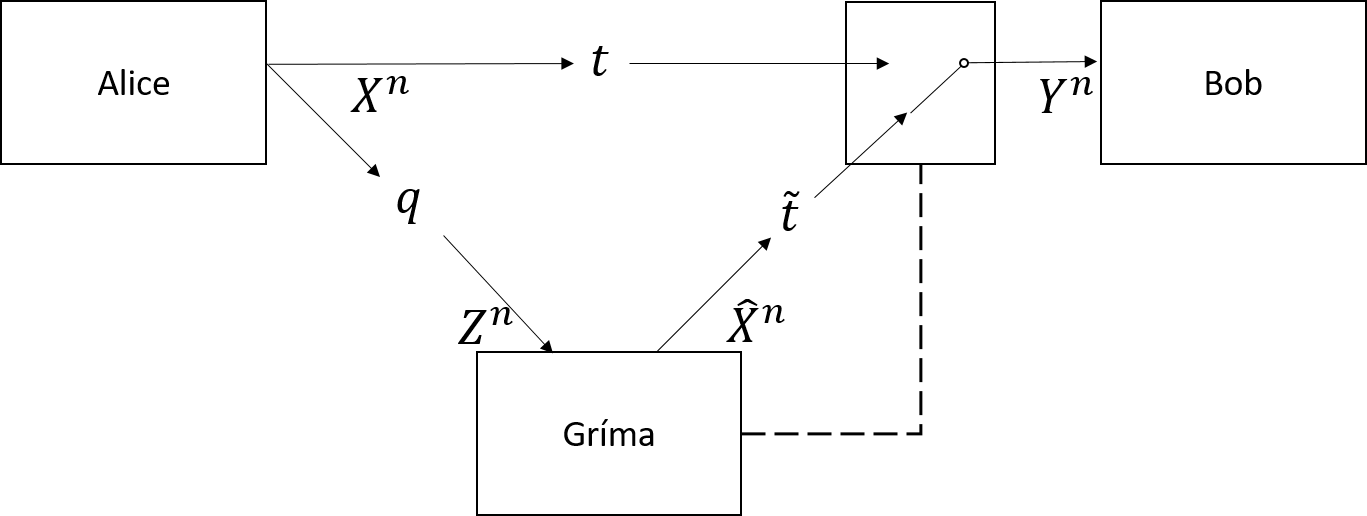}
\caption{Channel Model}
\label{fig:model}
\end{figure}

Our authentication model consists of three parties.
Alice, a legitimate transmitter, wishes to authenticate her communications with Bob, a legitimate receiver, over a discrete memoryless channel in the presence of Gr{\'i}ma, a malicious adversary.
Gr{\'i}ma has the ability to intercept Alice's message and send his own to Bob via a noiseless channel.
His goal is to have Bob accept his messages as if they were from Alice.
To aid in authentication, Alice and Bob share a secret key $ K $ which is distributed uniformly over $ \mcf{K} \defn \set{1, \dots, 2^{n \kappa}} $.

When Alice wishes to communicate, she jointly encodes a message $ M $, distributed uniformly over $ \mcf{M} \defn \set{ 1, \dots, 2^{n r}} $, and the key $ K $, as codeword $ X^n $.
The distribution of $ X^n $ is defined by the encoder $ f \in \mcf{P}(\mcf{X}^n|\mcf{M} ,\mcf{K}) $, where $ \mcf{P}(\mcf{X}^n|\mcf{M}, \mcf{K}) $ is the set of all probability distributions over $ \mcf{X}^n $ conditioned on $ \mcf{M}\times \mcf{K} $.
Alice then transmits $ X^n $ to both Bob and Gr{\'i}ma.
The three parties are connected via a \emph{discrete memoryless-adversarial substitution channel} (DM-ASC) which consists of three discrete memoryless channels, $(t,q,\tilde t)$, and a Gr{\'i}ma-controlled switch.
The triple represents the channels from Alice to Bob, Alice to Gr{\'i}ma, and Gr{\'i}ma to Bob, respectively, while the switch controls Bob's observations.

Note that for simplicity, we use the triple $ (t,q,\tilde t) $ instead of the formal septuple $ (\mcf{X}, \mcf{Y},\mcf{Z},\mcf{\tilde X},t,q,\tilde t) $, assuming that these values specify $ \mcf{X},~\mcf{Y},~\mcf{Z},~\mcf{\tilde X} $ by their non-zero indices.
Furthermore, we will assume for the remainder that $ \mcf{X},~\mcf{Y},~\mcf{Z},~\mcf{\tilde X} $ are all discrete and finite.
The channel is depicted in Figure \ref{fig:model}.

When the switch is open, Bob will receive Alice's transmission over $ t $.
In other words, $Y^n|X^n$ will be distributed according to $t^{\otimes n}(y^n|x^n) := \prod_{i=1}^n t(y_i|x_i)$ where $t \in \mcf{P}(\mcf{Y}|\mcf{X})$.
When the switch is closed, Gr{\'i}ma first obtains $Z^n|X^n$, then determines $\hat X^n$ and transmits to Bob.
We only consider $ (t,q,1) $ in which the channel from Gr{\'i}ma to Bob is noiseless, i.e. $ \mcf{\tilde X} = \mcf{Y} $, as in \cite{lai2009authentication,gungor2016basic}.
Thus, $Y^n|X^n$ will be distributed according to 
\[
\sum_{z^n} \psi(y^n|z^n) q^{\otimes n}(z^n|x^n),
\]
where $ q\in \mcf{P}(\mcf{Z}|\mcf{X}) $, and $ \psi \in \mcf{P}(\mcf{Y}^n|\mcf{Z}^n) $.
Gr{\'i}ma is free to choose any attack strategy, $ \psi $, including ones modeled after the standard impersonation and substitution attacks.
Regardless of the switch's position, Bob receives $Y^n$ and either makes an estimate of the message, $ M^* $, or declares an intrusion, $ \mbf{!} $, which is determined by a decoder $ \varphi \in \mcf{P}(\mcf{M}\cup \set{\mbf{!}} |\mcf{Y}^n, \mcf{K}) $.
%Gr{\'i}ma, meanwhile, receives $ Z^n $ and decides whether to attack.

\subsection{Performance Metrics}\label{sec:metrics}
Before presenting the performance metrics, we define an authentication code.
\begin{define}\label{def:code}
A \emph{code} is any pair $(f,\varphi)$, where $f \in \mcf{P}(\mcf{X}^n |\mcf{M} ,  \mcf{K})$ and $\varphi \in \mcf{P}(\mcf{M} \cup \set{\mbf{!}} |\mcf{Y}^n , \mcf{K})$.
The \emph{rate} of $(f,\varphi)$ is $n^{-1} \logt \abs{\mcf{M}}$, the \emph{block-length} of $(f,\varphi)$ is $n$, and the \emph{key requirement} of $(f,\varphi)$ is $n^{-1} \logt \abs{\mcf{K}}$.
\end{define}
The performance of the code is measured in two ways, reliability and type I error.
Reliability is measured by the average probability of error over all keys and messages at Bob, that is
\begin{equation}\label{eq:int:rel}
\varepsilon_{f,\varphi} \defn \abs{\mcf{K}}^{-1} \abs{\mcf{M}}^{-1} \sum_{m,k} \varepsilon_{f,\varphi}(m,k) < \epsilon ,
\end{equation}
where $ \epsilon \in (0,1) $ is a chosen constraint and
\begin{equation}
\varepsilon_{f,\varphi}(m,k) \defn 1-\sum_{x^n, y^n } \varphi(m|y^n,k)  t^{\otimes n} (y^n|x^n)  f(x^n|m,k).
\end{equation}
Type I error refers to the fact that authenticating is equivalent to a binary hypothesis test where the null hypothesis is an intrusion and the alternate hypothesis is authenticity.
Therefore, a good code limits the average type I error by
\begin{equation}\label{eq:int:at1e}
\omega_{f,\varphi} \defn \max_{\psi \in \mcf{P}(\mcf{Y}^n|\mcf{Z}^n)} E_{Z^n,M,K} \left[ \omega_{f,\varphi}(\psi,z^n,m,k) \right] \leq 2^{-na},
\end{equation}
where
\begin{equation}
\omega_{f,\varphi}(\psi,z^n,m,k) \defn \sum_{y^n} \psi(y^n|z^n) \varphi(\mcf{M} - \set{m} |y^n,k ).
\end{equation}

\begin{define}\label{def:average code}
A code $(f,\varphi)$ is called an $(r,\alpha,\kappa,\epsilon,n)$-\emph{average authentication} (AA) code for DM-ASC$(t,q,1)$ if the block-length is $n$, the rate at least $r$, the key requirement at most $\kappa$, it is reliable in that $\varepsilon_{f,\varphi}< \epsilon$ and it satisfies the average authenticity requirement:
\begin{align}\label{eq:average_auth}
\omega_{f,\varphi}  < 2^{-n \alpha}.
\end{align}
\end{define}
Our study aims to determine what types of codes are possible in the following sense.
\begin{define}
A triple $(a,b,c)$ is said to be \emph{achievable} for the DM-ASC $(t,q,1)$ if there exist a sequence of $\set{(r_i,\alpha_i,\kappa_i,\epsilon_i,i)}_{i=1}^\infty $-AA codes $(f_i,\varphi_i)$ such that 
\[
\lim_{i \rightarrow \infty} \abs{ (r_i,\alpha_i,\kappa_i,\epsilon_i,i) - (a,b,c,0,i)}_2 \rightarrow 0.
\] The \emph{average authentication region} (AAR) is then 
\begin{align}
\mcf{C}_{A}&(t,q,1) \nonumber\\
&\defn \set{(a,b,c) : (a,b,c) \text{ is achievable for DM-ASC}(t,q,1)  }.
\end{align} 
\end{define}	

\section{Background}\label{sec:background}
Before presenting the inner bound for the average authentication region, we review existing schemes.
First, we review Lai's \cite{lai2009authentication} strategy and frame it in terms of information metrics for ease of comparison.
Next, we examine Simmons' \cite{Auth} strategy for the noiseless channel and transform Gungor and Koksal's \cite{gungor2016basic} inner bound into our terms.

\subsection{Lai's Strategy}\label{sec:lai_review}

In \cite{lai2009authentication}, Lai et. al. propose essentially using a code designed for a wire-tapper channel, and sending the key as part of the message. The specific code they proposed is optimal for their limited scenario (key requirement $\rightarrow 0$), but in light of the forthcoming discussion, it is not optimal in ours.

Recognizing that the essence of the construction is to transmit two independent messages (the message itself and the key), with one subject to a secrecy constraint, the most logical coding scheme is a special class of codes for the \emph{discrete memoryless broadcast channel with confidential communications (t,q,)} (DM-BCC$ (t,q) $). While we are the first to notice and use this specific construction for the purpose of authentication, we refer to this as Lai's strategy. Before continuing we discuss the DM-BCC.

The achievable rate region of the DM-BCC was first derived by Csisz{\'a}r and K{\"o}rner in \cite{csiszar1978broadcast} and later refined in \cite[Chapter~17]{CK}.
In said model, there exist three messages that Alice wishes to send, a common message, $ m_0 \in \mcf{M}_0 \defn \set{1,\dots,2^{nr_0}} $, that is to be decoded by both Bob and Gr{\'i}ma, a private message, $ m_s \in \mcf{M}_s \defn \set{1, \dots, 2^{nr_s}} $, that is to be decoded by Bob and kept secret from Gr{\'i}ma, and finally a message, $ m_1 \in \mcf{M}_1\defn \set{1,\dots,2^{nr_1}} $, to be decoded by only Bob, but without a secrecy constraint.
Secrecy in this context is indicated by
\[
\msf{I}( p_{Z^n|M_s},p_{M_s} ) \leq \epsilon_n,
\]
where $\epsilon_n \rightarrow 0$ as $n\rightarrow \infty$. Meaning that the information gained about $ M_s $ from Gr{\'i}ma's observations asymptotically vanishes.
All messages have reliability constraints for their intended recipients.
The three messages are jointly coded as $ X^n $ and sent through the channel where Bob observes $ Y^n | X^n $, which is distributed as $ t^{\otimes n}(y^n|x^n) $ while Gr{\'i}ma observes, $ Z^n | X^n $, distributed as $ q^{\otimes n}(z^n|x^n) $.

The triple $(r_0,r_s,r_1)$ is achievable for the DM-BCC$ (t,q)$ if
\begin{align*}
r_0 + r_s + r_1 &\leq \msf{I}(t\rho ,\sigma |\tau ) + \min \left( \msf{I}(t\rho\sigma,\tau), \msf{I}(q\rho\sigma,\tau) \right)\\
r_s &\leq  \msf{I}(t\rho,\sigma|\tau)- \msf{I}(q\rho,\sigma|\tau)  \\
r_0 &\leq \min \left( \msf{I}(t\rho\sigma,\tau), \msf{I}(q\rho\sigma,\tau)\right) ,
\end{align*}
for some $\rho \in \mcf{P}(\mcf{X}|\mcf{U})$, $\sigma \in \mcf{P}(\mcf{U} \gg \mcf{W})$ and $\tau \in \mcf{P}(\mcf{W})$, and sets $\mcf{U}$ and $\mcf{W}$ such that $\abs{\mcf{U}} = (\abs{\mcf{X}}+1)(\abs{\mcf{X}}+3)$ and $\abs{\mcf{W}} = \abs{\mcf{X}}+3$.
It can be seen here that secrecy is only possible when the channel from Alice to Gr{\'i}ma's is not less noisy than the channel from Alice to Bob.

Lai's strategy attains authentication capabilities by implementing the coding scheme for the DM-BCC$ (t,q) $ in which Alice's message is sent as $ M_1 $ and the key is sent as $ M_s $ while $ M_0=\emptyset $.
If message rates are chosen within the achievable region above, Bob will decode the message reliably, satisfying the reliability constraint of an authentication code.
Additionally, since the key is also reliably decoded and each $ y^n $ corresponds to only one $ k $, he can declare authenticity when $ \hat k = k $.
The security constraint on $ M_s=K $ reduces the information about the key that is leaked to Gr{\'i}ma; the analysis of our work will determine the degree of effectiveness.

%The achievable region for Lai's strategy is
%\begin{align*}
%r + \alpha  &\leq \msf{I}(t\rho ,\sigma |\tau ) + \min \left( \msf{I}(t\rho\sigma,\tau), \msf{I}(q\rho\sigma,\tau) \right)\\
%\alpha &\leq  \msf{I}(t\rho,\sigma|\tau)- \msf{I}(q\rho,\sigma|\tau)  \\
%\alpha - \kappa &\leq 0.
%\end{align*}
As stated before, non-zero rates are only possible when $ t $ is less noisy than $ q $, i.e. when $ \msf{I}(t\rho,\sigma|\tau) > \msf{I}(q\rho,\sigma|\tau) $.
To solve this issue, we return to Simmons' strategy for the noiseless case. 

\subsection{Simmon's Strategy}
Simmons' authentication scheme \cite{Auth} for noiseless channels breaks down the problem into protecting against two different attacks, i.e., an impostor formerly referred to as ``impersonation'' attack and a substitution attack.
The attacks differ in that in the former, Gr{\'i}ma attacks without first observing one of Alice's transmissions, while in the latter, Gr{\'i}ma does.
In the strategy, the code is created by independently and randomly choosing $ |\mcf{K}|=2^{n\kappa} $ not necessarily unique subsets of $ \mcf{M} $, each denoted as $ \mcf{M}(k) \subset \mcf{M} $.
The size of each subset is $ |\mcf{M}(k)|=2^{-n\kappa/2} |\mcf{M}| $ where each element $ m \in \mcf{M}(k) $ corresponds to a single message $ \tilde m \in \mcf{\tilde M} \defn \set{1 , \dots, \abs{\mcf{M}}2^{-n \kappa/2}  } $.
Then, to communicate $ \tilde{m} $, Alice sends the associated $ m $ from the subset indexed by their shared key, $ k $.
Bob authenticates a message when the observed $ m $ is an element of the correct $ \mcf{M}(k) $.
The rate of communication in this scheme is $ n^{-1} \logt \abs{\mcf{\tilde M}} = n^{-1} \logt \abs{\mcf{M}} - \kappa/2 $. 

Since an observed $ m $ can be contained in multiple $ \mcf{M}(k) $, Gr{\'i}ma will be unable to immediately infer which key was used for authentication.
In order to launch a successful substitution attack, Gr{\'i}ma must choose an $ m' \neq m $ that is contained in the same $ \mcf{M}(k) $, however on average there is only $\abs{\mcf{K}} ( \abs{\mcf{M}(k)} /\abs{\mcf{M}})^2 =  1$ subset that contains both $ m $ and $ m' $.
Therefore, he must essentially guess the correct key to fool Bob which happens with probability $ 2^{-n \kappa/2} $ since there are, on average, $ \abs{\mcf{K}} ( \abs{\mcf{M}(k)} /\abs{\mcf{M}}) = 2^{n\kappa/2} $ subsets that contain $ m $.
In terms of an achievable rate region, this scheme can achieve the triple $(n^{-1} \logt \abs{\mcf{M}} - \kappa/2, \kappa/2, \kappa)$.
Simmons' strategy, together with Lai's strategy, forms the basis for our code.

\subsection{Gungor and Koksal's Bounds} 
Inner bounds for the average achievability region of a DM-ASC$(t,q,\tilde t)$, have been established by Gungor and Koksal~\cite{gungor2016basic}.
Specifically, their scheme splits Alice and Bob's shared key into two smaller keys, one for authentication ({\'a} la Lai's strategy) and one for secrecy.
These two keys are then used as the dimensions in a two dimensional dimensional binning process, where the codeword corresponding to the triple of messages and keys is chosen independently.
The independent choice over the secrecy key, though, leaks extraneous information since there is no need to differentiate between secrecy keys at the legitimate receiver.

In any case, the set of all achievable $(r,\alpha, \kappa)$ derived from their scheme is a subset of 
\begin{equation}\label{eq:reg_gungor}
(r,\alpha , \kappa) \in \bigcup_{\tilde \kappa \in \mathbb{R}_+ } \mcf{R}_{G}(\tilde \kappa)  
\end{equation}
where $	\mcf{R}_{G}(\tilde \kappa)$
\begin{equation}\label{eqn:gungorregion}
\defn \set{ r,\alpha,\kappa : \setlength\arraycolsep{.75pt}\begin{matrix*}[l]
r+ \kappa &\leq \msf{I}(t\rho,\tau) + \tilde \kappa  \\
\alpha - \kappa &\leq - \tilde \kappa \\
\alpha &\leq \min_{\nu \in \mcf{P}(\mcf{Z}|\mcf{U}) } \msf{L}_{G}(\nu,q\rho, t\rho, \tilde \kappa, \tau)
\end{matrix*}}
\end{equation}
and $ \msf{L}_{G}(\nu,q\rho, t\rho, \tilde \kappa, \tau) = \msf{D}(\nu||q\rho|\tau) + \left|  \tilde \kappa + \msf{I}(t\rho,\tau) - \msf{I}(\nu,\tau) \right|^+$. A proof of this can be found in \iftoggle{arxiv}{Appendix \ref{app:GungorRegion}.}{~\cite[Appendix A]{perazzone2018inner}.} %\ref{app:GungorRegion}]{perazzone2018inner}.}

\section{Authentication Capacity Region}\label{sec:thms}
We now present our main theorems and the inner bound of the average authentication region.
First, we present the minor contribution of characterizing the inner bound of the authentication region using Lai's strategy.

\begin{theorem}\label{thm:lai_strong} 
\begin{align}\label{eq:reg_lai:s}
\set{(r,\alpha,\kappa) :\setlength\arraycolsep{2pt} \begin{matrix*}[l]
r + \alpha &\leq \msf{I}(t\rho,\sigma\tau) \\
\alpha & \leq  \min_{\nu \in \mcf{P}(\mcf{Z}|\mcf{U})}  \msf{L}(\nu; t\rho,q\rho,\sigma,\tau) \\
\alpha & \leq  \msf{I}(t\rho,\sigma|\tau) \\
\alpha -  \kappa &\leq 0 
\end{matrix*} }& \nonumber \\
\subset \mcf{C}_{A}(t,q,1),&
\end{align}
where 
$ \msf{L}(\nu; t\rho,q\rho,\sigma,\tau) \defn  \msf{D}(\nu ||q\rho |\sigma\tau) + |\msf{I}(t\rho,\sigma|\tau) - \msf{I}(\nu , \sigma|\tau) + \left| \msf{I}(t\rho\sigma,\tau) - \msf{I}(\nu \sigma,\tau)  \right|^+ |^+$,
for all $\rho \in \mcf{P}(\mcf{X}|\mcf{U})$, $\sigma \in \mcf{P}(\mcf{U}\gg\mcf{W})$, and $\tau \in \mcf{P}(\mcf{W})$ where $\abs{\mcf{U}}$ and $\abs{\mcf{W}}$ are finite.
\end{theorem}
\begin{IEEEproof}
\iftoggle{arxiv}{See~Appendix \ref{app:laiavg}, along with the supporting code construction, message error analysis and type I error analysis in Appendix~\ref{app:cc1}.}
{See~\cite[Appendix C]{perazzone2018inner}, along with the supporting code construction, message error analysis and type I error analysis in~\cite[Appendix~F]{perazzone2018inner}.} %\ref{app:laiavg}]{perazzone2018inner} ref{app:cc1}]{perazzone2018inner}.}
\end{IEEEproof}
%The result of the theorem is as expected.
The type I error capabilities are limited by the capacity of the wire-tap channel and if the secrecy capacity is 0, then no authentication is possible.
We now extend Simmons' strategy and although it will only be applied to the triples from~Theorem~\ref{thm:lai_strong}, the associated code construction makes no such assumption on the genesis of the original code.
\begin{theorem}\label{thm:-t2}
If $(r,\alpha, \kappa) \in \mcf{C}_A$ then $(r-\beta,\alpha+\beta, \kappa+ 2\beta) \in \mcf{C}_{A}$, for all non-negative $\beta < r$.
\end{theorem}
\begin{IEEEproof}
See\iftoggle{arxiv}{ Appendix \ref{app:univcompos}}{~\cite[Appendix D]{perazzone2018inner}.}%\ref{app:univcompos}]{perazzone2018inner}}.
\end{IEEEproof}

Now to obtain our inner bound, we combine Theorems \ref{thm:lai_strong} and \ref{thm:-t2}.
\begin{theorem}\label{thm:ib}
\begin{align}\label{eqn:avgregion}
\set{(r,\alpha,\kappa) :\setlength\arraycolsep{2pt}\begin{matrix*}[l]
r + \alpha &\leq \msf{I}(t\rho,\sigma\tau) \\
2\alpha - \kappa  & \leq \min_{\nu \in \mcf{P}(\mcf{Z}|\mcf{U})}   \msf{L}(\nu;t\rho,q\rho,\sigma,\tau)\\
2\alpha - \kappa  & \leq \msf{I}(t\rho,\sigma|\tau) \\
\alpha -  \kappa &\leq 0 
\end{matrix*} }& \nonumber \\
\subset \mcf{C}_{A}(t,q,1),&
\end{align}
where $ \msf{L}(\nu; t\rho,q\rho,\sigma,\tau) \defn  \msf{D}(\nu ||q\rho |\sigma\tau) + |\msf{I}(t\rho,\sigma|\tau) - \msf{I}(\nu , \sigma|\tau) + \left| \msf{I}(t\rho\sigma,\tau) - \msf{I}(\nu \sigma,\tau)  \right|^+ |^+$,
for all distributions $\rho \in \mcf{P}(\mcf{X}|\mcf{U})$, $\sigma \in \mcf{P}(\mcf{U}\gg\mcf{W})$, and $\tau \in \mcf{P}(\mcf{W})$  and  $\abs{\mcf{U}}$ and $\abs{\mcf{W}}$ are finite.
\end{theorem}

\begin{IEEEproof}
The proof can be found in \iftoggle{arxiv}{Appendix \ref{app:avgregion}}{\cite[Appendix E]{perazzone2018inner}.} %\ref{app:avgregion}]{perazzone2018inner}}.
\end{IEEEproof}

This inner bound exhibits a trade-offs between rate, type I error, and key requirement in information theoretic terms.
It is apparent from the first condition that this scheme requires communication and authentication share the main channel's capacity.
As long as $ \min_{\nu \in \mcf{P}(\mcf{Z}|\mcf{U})}   \msf{L}(\nu;t\rho,q\rho,\sigma,\tau)  $ is non-zero, an increase in the length of the secret key provides a proportional increase in type I error.
Whereas when the condition is zero, an increase in $ \alpha $ requires twice the increase in $ \kappa $ as evident in Simmons' scheme.
%Since authentication is possible when the channel from Alice to Gr{\'i}ma is less noisy than the channel from Alice to Bob, it clearly improves Lai's region where this is not possible. 

Our scheme also improves over Gungor and Koksal's inner bound in this respect, since our scheme does not continue to unnecessarily leak information when Gr{\'i}ma's channel is less noisy than Bob's channel. Instead, in such a case, our scheme reverts to that of Simmon's, which is known to be optimal.

\section{Examples}\label{sec:examples}

To demonstrate that our inner bound outperforms Gungor and Koksal's inner, we provide a few examples and analyses.
While it is easy to see that our inner bound \eqref{eqn:avgregion} is larger than Lai's \eqref{eq:reg_lai:s} due to the addition of $ 2\alpha-\kappa $, we will provide an explicit example to show that~\eqref{eqn:avgregion} also improves upon Gungor's inner bound \eqref{eqn:gungorregion}.
For clarity, we will examine the case where $ t $ and $ q $ are binary symmetric channels (BSC) with transition probabilities $\lambda_t$ and $\lambda_q$ respectively.

%\subsection{BSC Analysis}
\pagebreak 
In a BSC, \eqref{eqn:avgregion} simplifies to 
\begin{align*}
r + \alpha &< \msf{I}(t,\sigma\tau)\\
2\alpha -\kappa &< \min_{\nu \in \mcf{P}(\mcf{Z}|\mcf{X})} \msf{L}^*(\nu; t,q,\sigma,\tau)\\
2\alpha - \kappa  & \leq \msf{I}(t,\sigma|\tau) \\
\alpha - \kappa &\le 0 ,
\end{align*}
where $ \sigma $ is now a distribution on $ X $ given $ W $ and $ \msf{L}^*(\nu; t,q,\sigma,\tau) = \msf{D}(\nu ||q |\sigma\tau) + |\msf{I}(t,\sigma|\tau) - \msf{I}(\nu , \sigma|\tau) + \left| \msf{I}(t\sigma,\tau) - \msf{I}(\nu \sigma,\tau)  \right|^+ |^+  $.
Meanwhile,~\eqref{eqn:gungorregion} simplifies to
\begin{equation}\label{eqn:gungorBSC}
\mcf{R}_{G}(\tilde \kappa)  \hspace{-3pt} \defn \hspace{-3pt}\set{ r,\alpha,\kappa : \setlength\arraycolsep{1pt}\begin{matrix*}[l] 
r+ \kappa &\leq \msf{I}(t,\tau) + \tilde \kappa  \\
\alpha - \kappa &\leq - \tilde \kappa \\
\alpha &\leq \min_{\nu \in \mcf{P}(\mcf{Y}|\mcf{X}) }  \msf{L}_{G}(\nu,q,t,\tilde \kappa,\tau)
\end{matrix*}},
\end{equation}
where $ \msf{L}_{G}(\nu,q,t,\tilde \kappa,\tau) =  \msf{D}(\nu||q|\tau) + \left|  \tilde \kappa + \msf{I}(t,\tau) - \msf{I}(\nu,\tau) \right|^+$.
%First, we will consider a case where the main channel $ t $ is less noisy than Gr{\'i}ma's channel $ q $, i.e. the main channel's transition probability, $ \lambda_t $, is less than $ \lambda_q $.
%In this case, our region and Gungor's are equivalent in which a greater range of rate and type I error trade-offs are provided when compared to Lai's strategy \textcolor{violet}{(this isn't true)}.
%We will show through explicit example that Gungor's region is a subset of the AAR.
% \iftoggle{arxiv}{}{In the interest of space, we will leave a more complete analysis in \cite{perazzone2018inner} and only present numerical examples here.}

\iftoggle{arxiv}{
\subsection{BSC Analysis}
The $ \sigma $ and $\tau$ that maximize the average region and all three constraints simultaneously is BSC(.5) and a uniform $ \tau $. 
Since the upper bound of the third condition is always larger than the upper bound of the second condition for this set of distributions, we only focus on the first two conditions.
For a less noisy main channel, the minimum $ \nu $ is always $ t $, whereas when the main channel is not less noisy, the minimum $ \nu $ is always $ q $, both regardless of Alice's choice of $ (\sigma,\tau) $.

When we consider the case where the adversary's channel is ``less noisy'' than the main channel, it is easy to see that Gungor's region is not better than the AAR.
With such a channel pair, the condition $ \min_{\nu \in \mcf{P}(\mcf{Z}|\mcf{X})} \msf{L}^*(\nu; t,q,\sigma,\tau) $ when evaluated at its minimum $ \nu =  q $, the bound is $ \msf{L}^*(q; t,q,\sigma,\tau) = 0 $.
This results in the condition $ \alpha < \kappa/2 $ for the average region.
In order for Gungor's strategy to be better, its upper bounds on $ \alpha $ must be greater than $ \kappa/2 $.
Examining the third condition in Gungor's region, it can been seen that in order to have a nonzero bound on $ \alpha $, $ \tilde \kappa > \msf{I}(q,\tau) - \msf{I}(t,\tau) $ since $ \nu=q $ is a valid choice.
Then, for both conditions to be greater than $ \kappa/2 $ we must have
\begin{align*}
&\tilde \kappa < \kappa/2 ~\\
&\tilde \kappa + \msf{I}(t,\tau) - \msf{I}(q,\tau) > \kappa/2.
\end{align*}
These two conditions, however, cannot occur simultaneously since they imply that $ \tilde \kappa + \msf{I}(t,\tau) - \msf{I}(q,\tau) > \tilde \kappa $ which is only valid when $ \msf{I}(t,\tau) > \msf{I}(q,\tau) $.
This is only satisfied when $ t $ is less noisy than $ q $ which contradicts our original assumption on this pair of channels.
Therefore, given the same $ \kappa $ and assumption of a less noisy adversary channel, Gungor's scheme cannot achieve any $ \alpha $ greater than that of our scheme, which implies that their region is contained in ours.
We will show through example that their region is a proper subset since there are instances of $ (r,\alpha,\kappa) $ that are contained in the AAR, but not in theirs.
}

\subsection{BSC Examples}

% \iftoggle{arxiv}
% {
% \begin{figure}
% 	\centering
% 	\includegraphics[scale=.4]{figures/AAvsGungorEqual}
% 	\caption{Given $ r $ and $ \kappa $, the AAR is equivalent to Gungor's region for a constant adversarial channel.}
% 	\label{fig:regionCompareEqual}
% \end{figure}
% }
% {
\begin{figure}
	\centering
	\includegraphics[scale=.35]{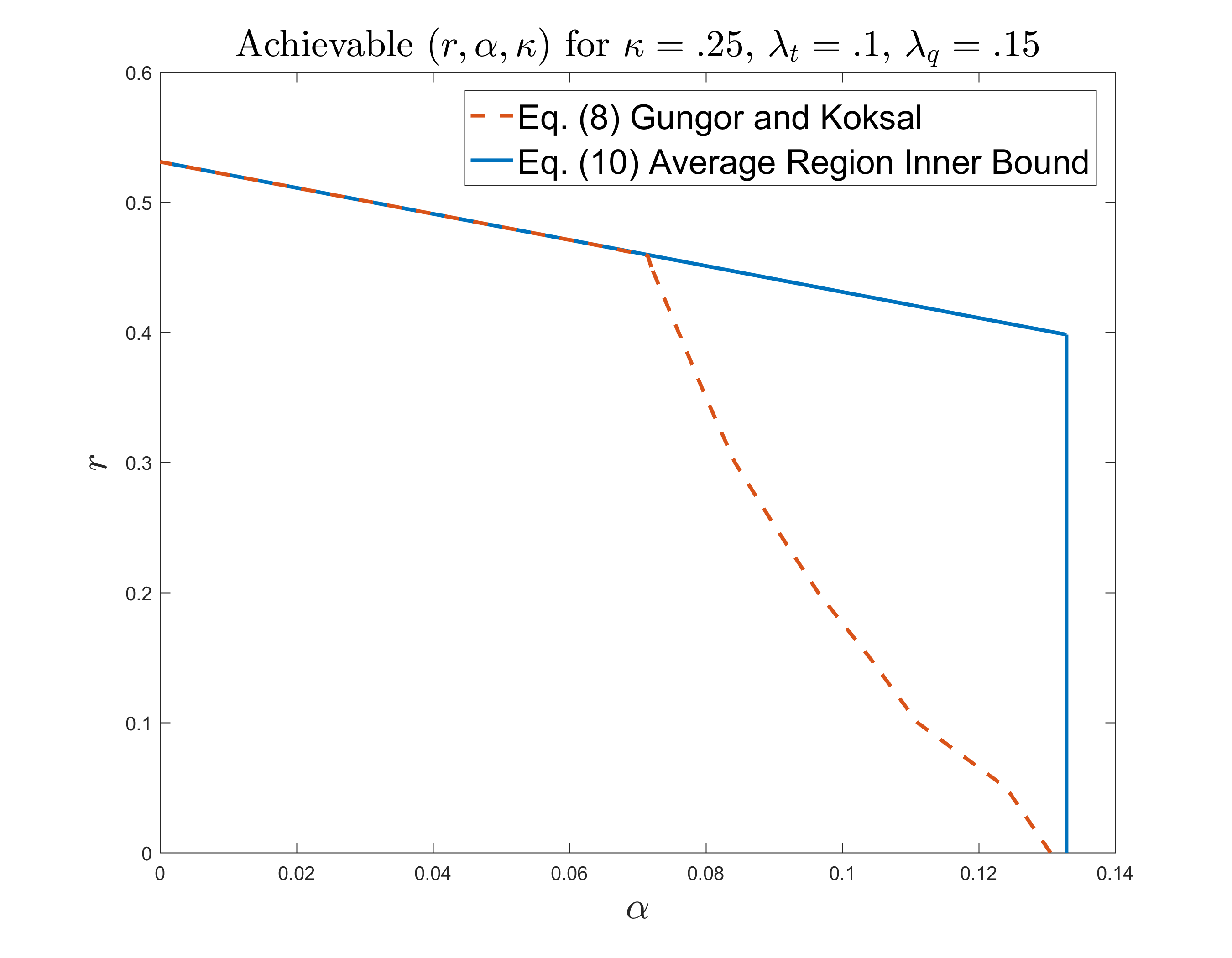}
	\caption{AAR outperforms Gungor's inner bound when $ \alpha $ is large.}
	\label{fig:regionCompareEqual}
\end{figure}
% }

\begin{figure}
  \centering
  \includegraphics[scale=.35]{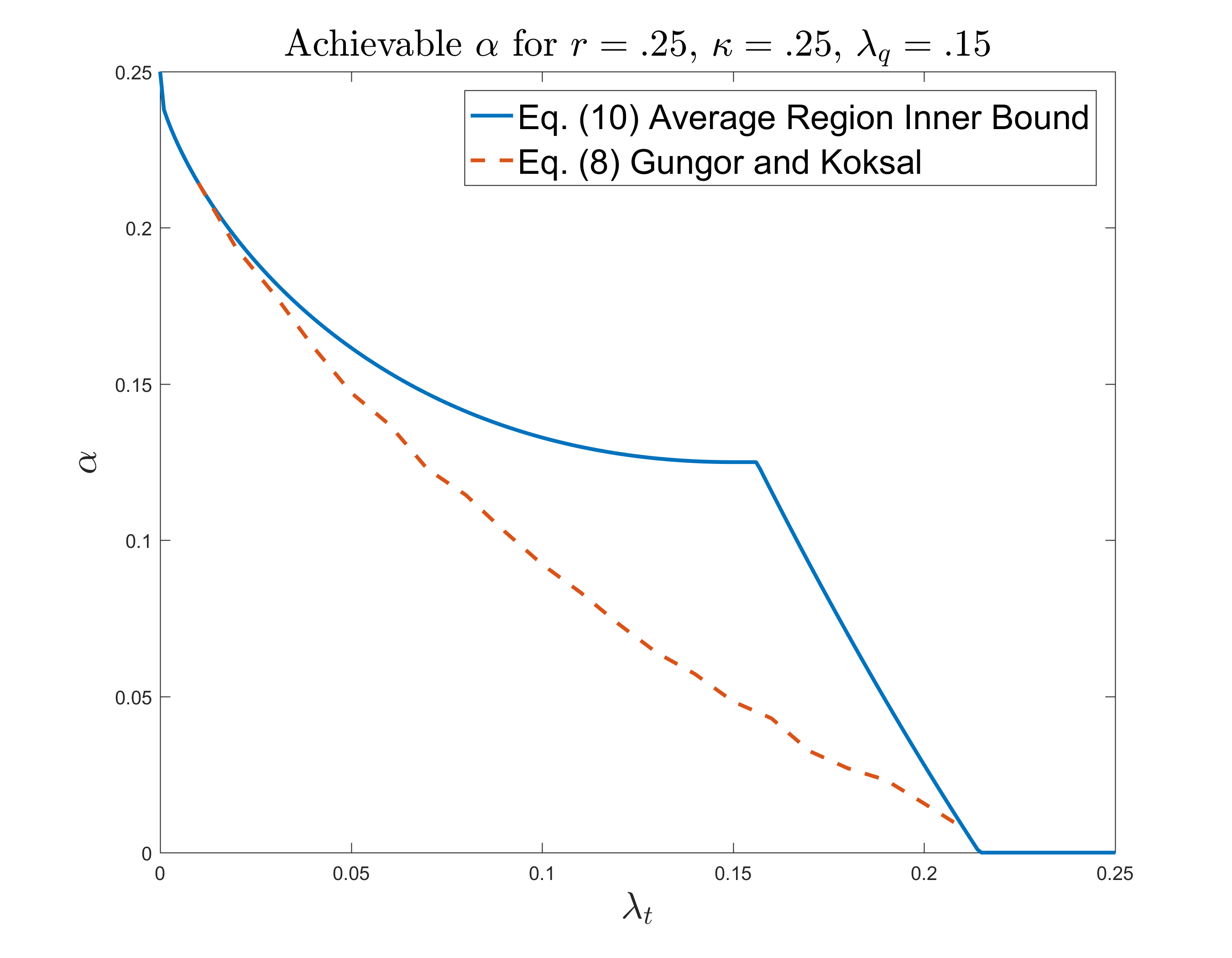}
  \caption{Given an $ r $ and $ \kappa $ pair, AAR achieves a greater range of $ \alpha $ for a constant adversarial channel ($ \lambda_t $ is the transition probability of channel $ t $).}
  \label{fig:regionCompare}
\end{figure}

First, we consider a case when the main channel is less noisy than Gr{\'i}ma's channel, where in specific $ \lambda_t=.1 $ and $ \lambda_q=.15 $.
The trade off between the rate and the authentication, given a fixed key rate, for both \eqref{eqn:avgregion} and~\eqref{eqn:gungorregion} is plotted in Figure~\ref{fig:regionCompareEqual}.
Note the equivalence of the two regions for small $ \alpha $.
As $\alpha$ increases, though, \eqref{eqn:avgregion} becomes strictly larger than~\eqref{eqn:gungorregion}.
While \eqref{eqn:avgregion} obtains a constant value for $r+\alpha$, which is equal to the capacity of $t$, approximately $.531$,~\eqref{eqn:gungorregion} struggles due to the inefficiency of their coding scheme.
This aligns with intuition, as \eqref{eqn:avgregion} uses the channel capacity for authenticity until the secrecy capacity is exhausted, and then switches to Simmons' scheme to further the authentication exponent.

Next, in Figure \ref{fig:regionCompare}, the rate, key requirement, and adversarial channel are held constant while the maximum possible $ \alpha $ achievable via~\eqref{eqn:avgregion} and~\eqref{eqn:gungorregion} is computed for a range of main channel transition probabilities, $ \lambda_t $.   
Both schemes have a dramatic performance decrease when the main channel becomes worse than the adversarial channel.
Still \eqref{eqn:avgregion} is generally larger than \eqref{eqn:gungorregion} for many possible main channels.
It should be noted the point where $\alpha = 0$ is exactly the point where the capacity of the channel equals $.25$, in other words both schemes are using all of the channels capacity simply to provide reliable communications.

\bibliographystyle{support/IEEEtran}
\bibliography{support/myrefs}

%	\section{Conclusion}
%	We have characterized the inner bound of the achievable region for the DM-ASC$ (t,q) $ under a reliability and average type I error constraint by combining \textcolor{orange}{the strategies of Lai and Simmons}.
%	The breakdown of information theoretic terms allows for easy comparison of the trade-offs between rate, type I error exponent, and key requirement.
%	We showed that our inner bound is an improvement of Lai et. al.'s region and Gungor and Koksal's inner bound since it can provide authentication when the adversarial channel is less noisy than the main channel.
%	
%	\textcolor{blue}{The conclusion can be cut for space considerations.}
\iftoggle{arxiv}{
\onecolumn
\appendix

\subsection{Proof of Gungor and Koksal's Region}\label{app:GungorRegion}
In order to make a fair comparison to our improved bound, it should be first noted that Gungor and Koksal's bound as they presented is incorrect. Specifically, the decoder they define on~\cite[Pg.~4535]{gungor2016basic}, is not the one they use for subsequent analysis. This is problematic as the mistake actually lets their scheme reliably transmit information at rates above channel capacity! After correcting this error~\cite[Theorem~1]{gungor2016basic} shows that to every $\rho \in \mcf{P}(\mcf{Y}|\mcf{U})$, $\tau \in \mcf{P}(\mcf{U})$, $(\kappa_1, \kappa_2) \in \mathbb{R}_+^2$ and large enough $n$ there exists a $(r,\alpha,\kappa_1 + \kappa_2, \varepsilon_{f,\varphi},n)$-AA code where
\begin{align*}
&\varepsilon_{f,\varphi} \dot \leq 2^{-n\left| \msf{I}(t\rho,\tau) - r - \kappa_1 \right|^+ } \\
&\alpha \dot = \min_{\nu \in \mcf{P}(\mcf{Z}|\mcf{U}) } \msf{D}(\nu||q\rho|\tau) + \min( \kappa_1 , \left| r + \kappa_1 + \kappa_2 - \msf{I}(\nu,\tau) \right|^+ )) .
\end{align*}
In this case, our declaration of intrusion, $ \mbf{!} $, is equivalent to their definition of an erasure, $ \alpha $, in \cite{gungor2016basic}.
Thus, $ \varepsilon_{f,\varphi} $ is less than the superposition of their undetected error and erasure bounds.
Note that, since the probability of a successful impostor attack is always less than that of a substitution attack, we only consider the latter.

These bounds imply that their inner bound is equal to~\eqref{eq:reg_gungor}, which follows by observing that all $(r,\alpha,\kappa)$ such that 
\begin{align*}
r   &\leq \msf{I}(t\rho,\tau)  - \kappa_1 \\
\kappa&=\kappa_1 + \kappa_2 \\
\alpha&\leq \kappa_1 \\
\alpha &\leq \min_{\nu \in \mcf{P}(\mcf{Z}|\mcf{U}) } \msf{D}(\nu||q\rho|\tau) + \left|  r + \kappa_1 + \kappa_2  - \msf{I}(\nu,\tau) \right|^+
\end{align*}
cause $\epsilon_{f,\varphi} \rightarrow 0$, hence are achievable, and that the last bound can be replaced by 
\[
\min_{\nu \in \mcf{P}(\mcf{Y}|\mcf{U}) } \msf{D}(\nu||q\rho|\tau) + \left|  \kappa_2 + \msf{I}(t\rho,\tau)  - \msf{I}(\nu,\tau) \right|^+ .
\]
Equation~\eqref{eq:reg_gungor} comes from applying Fourier Motzkin elimination and removing redundant equations, where $\tilde \kappa = k_2$.

\subsection{Code Construction}

The method of types will be used heavily in the code construction proofs related to Lai's strategy.
In order to facilitate the proofs, we now list a few important properties, but leave derivations to references~\cite[Chapter~2]{CK} or~\cite[Chapter~11]{CT}.
For $(y^n,x^n) \in \mcf{Y}^n\times \mcf{X}^n$, the important properties are, 
\begin{align}
t^{\otimes n}(y^n|x^n) &= 2^{-n \left( \msf{H}(\nu |\rho ) + \msf{D}(\nu || t |\rho ) \right)},\\
\abs{\mcf{T}^n_{\mu}(x^n)} &\dot = 2^{n \msf{H}(\mu|\rho) + O(\logt n)} ,\\
t^{\otimes n}(\mcf{T}^n_{\mu}(x^n)|x^n) &\dot = 2^{-n  \msf{D}( \mu || t |\rho ) + O(\logt n)},\\
\abs{\mcf{P}_n(\mcf{Y},\mcf{X})}  &\dot = n^{O(1)},
\end{align}
where $\nu = p_{y^n|x^n}$ and $\rho = p_{x^n}$.
All orders are determined solely by the cardinalities of support sets.
In addition to these well known properties, we will need the following lemma.
\begin{lemma}\label{lem:unique}
Let $\nu \in \mcf{P}(\mcf{Z}|\mcf{U})$, $\sigma \in \mcf{P}(\mcf{U} \gg \mcf{W})$ and $\tau \in \mcf{P}(\mcf{W})$. If $z^n \in \mcf{T}_{\nu}(u^n)$ and $u^n \in \mcf{T}_{\sigma}(w^n)$, then $z^n \in \mcf{T}_{\nu\sigma}(w^n)$. 
\end{lemma}
\begin{IEEEproof}
See Appendix~\ref{app:unique}.
\end{IEEEproof}
This is necessary since generally $z^n \notin \mcf{T}_{\nu\sigma}(w^n)$ when $z^n \in \mcf{T}_{\nu}(u^n)$ and $u^n \in \mcf{T}_{\sigma}(w^n)$.
Indeed, consider the case where $w^n = (0,0,1,1)$, $u^n = (1,0,1,0)$ and $z^n = (0,0,1,1)$.
Clearly $\nu(0|0) = \nu(1|0) = \nu(0|1) = \nu (1|1) = 1/2$ and $\sigma = \nu = \nu \sigma$.
But $p_{z^n|w}(1|1) = p_{z^n|w}(0|0) = 1$, and thus $p_{z^n|w^n} \neq \nu \sigma$.

We now turn to random coding arguments which will be used in the construction of the code.
In Lai's strategy, for example, a large number of $w^n$ will be independently selected uniformly at random from $\mcf{T}^n_{\tau}$, where $\tau \in \mcf{P}_n(\mcf{W})$, and then for each $w^n$ another large set of $u^n$ will be independently selected uniformly at random from $\mcf{T}^n_{\sigma}(w^n)$, where $\sigma \in \mcf{P}_n(\mcf{U} \gg \mcf{W})$.
In order to determine how many values of $u^n$ are chosen such that $y^n \in \mcf{T}^n_{\mu}(u^n)$ or $z^n \in \mcf{T}^n_{\nu}(u^n)$, we first need to introduce the following Lemma, which is in fact a minor result from Csisz{\'a}r and K{\"o}rner~\cite[Lemma~10.1]{CK}.
\begin{lemma}\textbf{\cite[Minor~result~from~Lemma~10.1]{CK}}\label{lem:obvious}
Let $U^n$ be uniformly distributed over $\mcf{T}^n_{\sigma}(w^n)$, for $w^n \in \mcf{T}^n_{\tau}$. If $\sigma \in \mcf{P}_{n}(\mcf{U}\gg \mcf{W})$ and $\tau \in \mcf{P}_{n}(\mcf{W})$, then
\[
\Pr \left( y^n \in \mcf{T}^n_{\mu}(U^n) \right) \dot = 2^{-n \msf{I}(\mu,\sigma|\tau) + O(\logt n)}
\]
for any $y^n \in \mcf{T}_{\mu\sigma}(w^n)$.
\end{lemma}
\begin{IEEEproof}
See Appendix~\ref{app:obvious}.
\end{IEEEproof}

The result is useful for both the reliability and security aspects of code construction analysis.
In particular, for reliability it is desirable to have the probability of choosing a $U^n$ such that $y^n \in \mcf{T}^n_{\mu}(U^n)$ be very small, so that a $u^n$ satisfying that relationship is unique.
On the other hand, it would be desirable to have a large probability of $z^n \in \mcf{T}^n_{\nu}(U^n)$ so that Gr{\'i}ma will have a large number of equally likely codewords associated with his observation.
The following lemma helps to meet this need.
\begin{lemma}\label{lem:ck} \textbf{(\cite[Lemma~17.9]{CK})} The probability that in $k$ independent trials an event of probability $q$ occurs less/more than $\alpha q k$ times, according as $\alpha \lessgtr 1$, is bounded above by $e^{-c(\alpha)qk}$ where $c(\alpha) = \alpha \ln \alpha - \alpha + 1$.
\end{lemma}
In determining the number of sequences of $u^n$ chosen such that $z^n \in \mcf{T}^n_{\nu}(u^n)$, each independent selection of a sequence can be viewed as a trial.
Lemma can be applied effectively regardless of if $qk \lessgtr 1$, although the method by which it should be applied differs.
To streamline the analysis, the following corollary of Lemma~\ref{lem:ck} will be used instead.
\begin{cor}\label{cor:ck} Fix $(\beta , \rho ) \in \mathbb{R}_{+}^2$ as well as functions $f,g : \mathbb{N} \rightarrow \mathbb{R}$. Suppose $V^{2^{n \beta + f(n) }} \defn (V_1, \dots, V_{2^{n\beta + f(n)}})$ are independent random variables that take $1$ with probability $2^{-n\rho - g(n)} $ and $0$ otherwise. Let
\[
\mcf{\hat V}_{\delta} \defn \set{ v^{2^{n \beta + f(n) }}  :  \left \lfloor 2^{-n\left| \beta - \rho\right|^+ - \left| f(n) - g(n) \right| - \delta  } \right \rfloor < \sum_{i} \idc{v_i = 1} \leq  2^{-n\left| \beta - \rho\right|^+ + \left| f(n) - g(n) \right| + \delta } }.
\]
For all $\beta,~\rho,~f$ and $g$
\begin{equation}
\Pr \left( V^{2^{n \beta + f(n) }} \in \mcf{\hat V}_{2 \logt ne } \right) \geq 1 - 2e^{-\frac{n^2}{4}}.
\end{equation}
Furthermore, if $n(\beta - \rho) + f(n) - g(n) > 2 \logt n$ then 
\begin{equation}
\Pr \left( V^{2^{n \beta + f(n) }} \in \mcf{\hat V}_{2 \logt e } \right) \geq 1 - 2e^{-\frac{n^2}{4}}.
\end{equation}
\end{cor}
\begin{IEEEproof}
See Appendix~\ref{app:ck}.
\end{IEEEproof}
Last, the following lemma will be required in the type I error analysis of our code. 
\begin{lemma}\label{lem:tbcut}
Let $\mcf{A}_1 , \dots , \mcf{A}_{k}$ be equal-sized subsets of some finite set $\mcf{X}$ (i.e, $\abs{\mcf{A}_1} = \abs{\mcf{A}_2} = \dots = \mcf{A}_{k}$). 
\[
\frac{k \abs{\mcf{A}_1} }{\max_{x \in \cup_{i} \mcf{A}_i} \abs{i : x \in \mcf{A}_i }  } \leq \abs{\cup_{i} \mcf{A}_i } < \frac{k \abs{\mcf{A}_1} }{\min_{x \in \cup_{i} \mcf{A}_i} \abs{i : x \in \mcf{A}_i }  }.
\]
\end{lemma}
\begin{IEEEproof}
See Appendix~\ref{app:tbcut}.
\end{IEEEproof}
\subsection{Proof of Theorem \ref{thm:lai_strong}}\label{app:laiavg} %Thm15

\begin{IEEEproof} 

First, fix a finite $\mcf{U}$ and $\mcf{W}$, as well as a $\rho \in \mcf{P}(\mcf{X}|\mcf{U})$, $\sigma \in \mcf{P}(\mcf{U}\gg\mcf{W})$, $\tau \in \mcf{P}(\mcf{W})$ such that
\[
\min_{\nu \in \mcf{P}(\mcf{Z}|\mcf{U})} \msf{L}(\nu;t\rho,q\rho,\sigma,\tau) > 0.
\]  
Also define
\begin{align*}
\tau_n &:= \argmin_{\tau' \in \mcf{P}_n(\mcf{W}) } \abs{\tau' - \tau}\\
\sigma_n &:= \argmin_{\sigma' \in \mcf{P}_n(\mcf{U}\gg\mcf{W}): \text{ if } \sigma(u|w) = 0 \text{ then } \sigma'(u|w) = 0} \abs{\sigma' - \sigma}\\
\Delta_{1,n} &:= \max_{\mu \in \mcf{P}(\mcf{Y}|\mcf{U})} \max_{n'>n}  \abs{\msf{D}(\mu || t \rho |\sigma_{n'}\tau_{n'}) - \msf{D}(\mu || t \rho |\sigma\tau) }\\
\Delta_{2,n} &:= \max_{\mu \in \mcf{P}(\mcf{Y}|\mcf{U})} \max_{n'>n}  \abs{\msf{I}(\mu,\sigma_{n'}|\tau_{n'}) - \msf{I}(\mu,\sigma|\tau) } \\
\Delta_{3,n} &:= \max_{\mu \in \mcf{P}(\mcf{Y}|\mcf{U})} \max_{n'>n}  \abs{\msf{I}(\mu,\sigma_{n'}\tau_{n'}) - \msf{I}(\mu,\sigma\tau) } \\
\Delta_{4,n} &:= \max_{\nu \in \mcf{P}(\mcf{Z}|\mcf{U})} \max_{n'>n}  \abs{\msf{D}(\nu || q \rho |\sigma_{n'}\tau_{n'}) - \msf{D}(\nu || q \rho |\sigma\tau) }\\
\Delta_{5,n} &:= \max_{\nu \in \mcf{P}(\mcf{Z}|\mcf{U})} \max_{n'>n}  \abs{\msf{I}(\nu,\sigma_{n'}|\tau_{n'}) - \msf{I}(\nu,\sigma|\tau) } \\
\Delta_{6,n} &:= \max_{\nu \in \mcf{P}(\mcf{Z}|\mcf{U})} \max_{n'>n}  \abs{\msf{I}(\nu,\sigma_{n'}\tau_{n'}) - \msf{I}(\nu,\sigma\tau) } \\
\Delta_{n} &:= \max_{i \in \set{1,\dots,6}} \Delta_{i,n},
\end{align*}
which will allow us to account for the code construction requiring empirical distributions for $\tau$ and $\sigma$. For this reason it is important to note that $\Delta_{n}$ goes to zero monotonically. Indeed this is because of the continuity of entropy since $\sigma_n \rightarrow \sigma$ and $\tau_n \rightarrow \tau$. 

First we will show that if $(r,\alpha,\kappa,\hat r, \tilde r , \tilde \kappa) \in \mcf{R}(\gamma_1,\gamma_2,\rho,\sigma,\tau)$, where
\begin{align}
\mcf{R}(\gamma_1,\gamma_2,\rho,\sigma,\tau) &:= \set{ (r,\alpha,\kappa,\hat r, \tilde r, \tilde \kappa)  : \begin{matrix*}[l] 
r - \tilde r - \hat r &\leq 0\\
\tilde \kappa - \kappa  &\leq 0 \\
\tilde r + \tilde \kappa &= \msf{I}(t\rho,\sigma|\tau) - \gamma_1 \\
\tilde r + \hat r + \tilde \kappa &= \msf{I}(t\rho,\sigma\tau) - \gamma_2 \\
\alpha - \tilde \kappa  &< 0 \\
\alpha &<    \min_{\nu \in \mcf{P}(\mcf{Z}|\mcf{U})} \msf{D}(\nu ||q\rho |\sigma\tau) + \left|\kappa_{\nu}^*(\gamma_1,\gamma_2,\rho,\sigma,\tau) \right|^+ 
%\alpha &< \min_{\nu \in \mcf{P}(\mcf{Z}|\mcf{U})} \msf{D}(\nu ||q\rho |\sigma\tau) + \left| \tilde \kappa - \left|\msf{I}(\nu,\sigma|\tau) - \tilde r -\left| \hat r - \msf{I}(\nu\sigma,\tau)  \right|^+  \right|^+ \right|^+ \label{eq:lai_strong:alpha}.
\end{matrix*}}
\end{align}
and
\[
\kappa_{\nu}^*(\gamma_1,\gamma_2,\rho,\sigma,\tau) = \msf{I}(t\rho,\sigma|\tau)  - \msf{I}(\nu,\sigma|\tau)  - \gamma_1  + \left| \msf{I}(t\rho\sigma,\tau) - \msf{I}(\nu\sigma,\tau) + \gamma_1 -\gamma_2  \right|^+,
\]
for some strictly positive $\gamma_1$ and $\gamma_2$, then $(r,\alpha, \kappa)$ is achievable. For large enough $n$, $r \leq \hat r + \tilde r$ and $\kappa \geq \tilde \kappa$, Appendix~\ref{app:cc1:cc},~\ref{app:cc1:mea}, and~\ref{app:cc1:t1e} demonstrate the existence of a $(r,\alpha_n,\kappa,\epsilon_n,n)$-AA code such that 
\begin{align}
\epsilon_{n}  &< \max_{\mu \in \mcf{P}(\mcf{Y}|\mcf{U})} 2^{- n\msf{D}(\mu||t\rho|\sigma_n\tau_n) + \chi \logt n} \max \left(  2^{n\left| \tilde r + \tilde \kappa -  \msf{I}(\mu,\sigma_n|\tau_n)\right|^-  } , 2^{n \left| \tilde r + \hat r + \tilde \kappa - \msf{I}(\mu,\sigma_n\tau_n)\right|^- }  \right) \label{eq:lais:me2} 
\end{align}
and 
\begin{equation}\label{eq:lais:t1e}
\alpha_n > \min_{\nu \in \mcf{P}(\mcf{Z}|\mcf{U})}  \left( \msf{D}(\nu ||q\rho |\sigma_n\tau_n) + \left| \tilde \kappa - \left|\msf{I}(\nu,\sigma_n|\tau_n) - \tilde r -\left| \hat r - \msf{I}(\nu\sigma_n,\tau_n)  \right|^+  \right|^+ \right|^+ \right) - \chi n^{-1} \logt n ,
\end{equation}
where $\chi$ is some positive constant only dependent on the cardinalities of $\mcf{Y},~\mcf{X},~\mcf{Z},~\mcf{U}$ and $\mcf{W}$. As we will show, if $(r,\alpha,\kappa,\hat r, \tilde r , \tilde \kappa) \in \mcf{R}(\gamma_1,\gamma_2,\rho,\sigma,\tau)$ then $\epsilon_n \rightarrow 0$ and $\alpha \leq \lim_{n \rightarrow \infty} \alpha_n$. Thus demonstrating that $(r,\alpha,\kappa)$ is achievable if $(r,\alpha,\kappa,\hat r, \tilde r , \tilde \kappa) \in \mcf{R}(\gamma_1,\gamma_2,\rho,\sigma,\tau)$ for some strictly positive $\gamma_1$ and $\gamma_2$. 

Starting with showing that if $(r,\alpha,\kappa,\hat r, \tilde r , \tilde \kappa) \in \mcf{R}(\gamma_1,\gamma_2,\rho,\sigma,\tau)$, then the RHS of Equation~\eqref{eq:lais:me2} decays to zero. First note that
\begin{align}
\epsilon_{n}  &< \max_{\mu \in \mcf{P}(\mcf{Y}|\mcf{U})} 2^{- n\msf{D}(\mu||t\rho|\sigma_n\tau_n) + \chi \logt n}  \max \left(  2^{n\left| \tilde r + \tilde \kappa -  \msf{I}(\mu,\sigma_n|\tau_n)\right|^-  } , 2^{n \left| \tilde r + \hat r + \tilde \kappa - \msf{I}(\mu,\sigma_n\tau_n)\right|^- }  \right) \notag\\
&< \max_{\mu \in \mcf{P}(\mcf{Y}|\mcf{U})} 2^{- n\msf{D}(\mu||t\rho|\sigma \tau ) + 2\Delta_n + \chi \logt n}  \max \left(  2^{n\left| \tilde r + \tilde \kappa -  \msf{I}(\mu,\sigma|\tau)\right|^-  } , 2^{n \left| \tilde r + \hat r + \tilde \kappa - \msf{I}(\mu,\sigma\tau)\right|^- }  \right) \label{eq:lais:eub1}\\
&= \max_{\mu \in \mcf{P}(\mcf{Y}|\mcf{U})} 2^{- n\msf{D}(\mu||t\rho|\sigma \tau ) + 2\Delta_n + \chi \logt n}  \max \left(  2^{n\left| -\gamma_1 + \msf{I}(t\rho,\sigma|\tau)  -  \msf{I}(\mu,\sigma|\tau)\right|^-  } , 2^{n \left| - \gamma_2 + \msf{I}(t\rho,\sigma\tau) - \msf{I}(\mu,\sigma\tau)\right|^- }  \right) ,\label{eq:lais:eub2} 
\end{align}
where~\eqref{eq:lais:eub1} is by the definition of $\Delta_n$, and~\eqref{eq:lais:eub2} is because $(r,\alpha,\kappa,\hat r, \tilde r, \tilde \kappa) \in \mcf{R}(\gamma_1,\gamma_2,\rho,\sigma,\tau)$. The $\mu$ that maximizes~\eqref{eq:lais:eub2} is independent of $n$. Observe that if the maximizing distribution, $\mu$, equals $t \rho$, then~\eqref{eq:lais:eub2} equals 
\begin{equation}
2^{-n (\min(\gamma_1,\gamma_2)) + 2n\Delta_n + \chi \logt n},
\end{equation}
which decays to zero since $\gamma_1$ and $\gamma_2$ are strictly positive while $\Delta_n \rightarrow 0$. On the other hand, if the maximizing distribution is not equal to $t \rho$ then~\eqref{eq:lais:eub2} is less than
\begin{equation}
2^{-n \msf{D}(\mu||t\rho|\sigma\tau) + 2n \Delta_n + \chi \logt n },
\end{equation}
which still decays to zero since $\msf{D}(\mu||t\rho|\sigma\tau)$ is strictly positive and independent of $n$. 

Next we show that $\alpha < \lim_{n \rightarrow \infty} \alpha_n$ for all $(r,\alpha,\kappa,\hat r, \tilde r, \tilde \kappa) \in \mcf{R}(\gamma_1,\gamma_2,\rho,\sigma,\tau)$. Similar to before, note
\begin{align}
\lim_{n \rightarrow \infty} \alpha_{n}  &= \min_{\nu \in \mcf{P}(\mcf{Z}|\mcf{U})}   \msf{D}(\nu ||q\rho |\sigma\tau) + \left| \tilde \kappa - \left|\msf{I}(\nu,\sigma|\tau) - \tilde r -\left| \hat r - \msf{I}(\nu\sigma,\tau)  \right|^+  \right|^+ \right|^+  \label{eq:lais:t1eub1}\\
&=\min_{\nu \in \mcf{P}(\mcf{Z}|\mcf{U})}   \msf{D}(\nu ||q\rho |\sigma\tau) + \left| \tilde \kappa - \left|\tilde \kappa - \kappa_{\nu}^*(\gamma_1,\gamma_2,\rho,\sigma,\tau)   \right|^+ \right|^+    \label{eq:lais:t1eub2}\\
&=\min_{\nu \in \mcf{P}(\mcf{Z}|\mcf{U})}   \msf{D}(\nu ||q\rho |\sigma\tau) + \min( \tilde \kappa, \abs{\kappa_{\nu}^*(\gamma_1,\gamma_2,\rho,\sigma,\tau)}^+)     \label{eq:lais:t1eub3}
\end{align}
where~\eqref{eq:lais:t1eub1} is because $\Delta_n$ and $\chi n^{-1}  \logt n$ go to zero, and~\eqref{eq:lais:t1eub2} is because $(r,\alpha,\kappa,\hat r, \tilde r, \tilde \kappa) \in \mcf{R}(\gamma_1,\gamma_2,\rho,\sigma,\tau)$. Let $\tilde \nu$ be the distribution in $\mcf{P}(\mcf{Z}|\mcf{U})$ which minimizes~\eqref{eq:lais:t1eub3}. Now\footnote{Although tangential to the proof, it should also be observed that  $\alpha < \min_{\nu} \msf{D}( \nu ||q\rho |\sigma\tau) + \min( \tilde \kappa, \abs{\kappa_{ \nu}^*(\gamma_1,\gamma_2,\rho,\sigma,\tau)}^+)$ implies $\alpha < \tilde \kappa$ and $\alpha< \min_{\nu} \msf{D}( \nu||q\rho|\sigma\tau) + \tilde \kappa_{\nu}^*(\gamma_1,\gamma_2,\rho,\sigma,\tau)$.} 
\begin{equation}
\alpha < \msf{D}(\tilde \nu ||q\rho |\sigma\tau) + \min( \tilde \kappa, \abs{\kappa_{\tilde \nu}^*(\gamma_1,\gamma_2,\rho,\sigma,\tau)}^+) = \lim_{n\rightarrow \infty} \alpha_n
\end{equation}
since 
\begin{equation}
\alpha < \tilde \kappa \leq \msf{D}(\tilde \nu||q\rho|\sigma\tau) + \tilde \kappa
\end{equation} 
and
\begin{equation}
\alpha < \msf{D}(\tilde \nu||q\rho|\sigma\tau) + \tilde \kappa_{\tilde \nu}^*(\gamma_1,\gamma_2,\rho,\sigma,\tau)
\end{equation} 
by the definition of $\mcf{R}(\gamma_1,\gamma_2,\rho,\sigma,\tau)$. 

Having shown $(r,\alpha,\kappa, \hat r , \tilde r, \tilde \kappa) \in \mcf{R}(\gamma_1,\gamma_2,\rho,\sigma,\tau)$ precipitates  $(r,\alpha,\kappa)$ being achievable, we begin the task of finding a simple characterization of all such triples in $\bigcup_{\gamma_1,\gamma_2,\rho,\sigma,\tau} \mcf{R}(\gamma_1,\gamma_2,\rho,\sigma,\tau)$. Let $\mcf{R}^*(\gamma_1,\gamma_2,\rho,\sigma,\tau)$ be the set of achievable $(r,\alpha,\kappa)$ via $\mcf{R}(\gamma_1,\gamma_2,\rho,\sigma,\tau)$, for which   
\begin{equation}
\mcf{R}^*(\gamma_1 , \gamma_2,\rho,\sigma,\tau) = \left\{(r,\alpha,\kappa): \begin{matrix*}[l] 
\alpha - \kappa &< 0\\
\alpha + r &< \msf{I}(t\rho,\sigma\tau) - \gamma_2 \\
\alpha &< \msf{I}(t\rho,\sigma|\tau)-\gamma_1 \\
\alpha &< \min_{\nu \in \mcf{P}(\mcf{Z}|\mcf{U}) } \msf{D}(\nu||q\rho|\sigma\tau) + \left|\kappa_{\nu}^*(\gamma_1,\gamma_2,\rho,\sigma,\tau) \right|^+
\end{matrix*}
\right\}
\end{equation}
by applying Fourier-Motzkin elimination\footnote{In order to obtain these equations, the additional requirements of $r,~\alpha,~\kappa,~\hat r,~\tilde r$ and $\tilde \kappa$ being non-negative are needed, these are equations are suppressed in the presentation.} to $\mcf{R}(\gamma_1,\gamma_2,\rho,\sigma,\tau)$. 

Next the $\gamma_1, \gamma_2$ terms can be removed since 
\begin{equation}\label{eq:lais:first_reduction}
\mcf{R}^*(\gamma_1,\gamma_2,\rho,\sigma ,\tau) \subseteq \mcf{R}^*(\gamma_1^*,\gamma_2^*,\rho,\sigma,\tau)
\end{equation}
for any $0 < \gamma_1^* \leq \gamma_1$ and $0 < \gamma_2^*\leq \gamma_2$.  Thus 
\begin{equation}
\cup_{\gamma_1,\gamma_2 } \mcf{R}^*(\gamma_1,\gamma_2,\rho,\sigma,\tau) = \mcf{R}^*_0(\rho,\sigma,\tau) := \lim_{(\gamma_1,\gamma_2) \rightarrow (0,0)} \mcf{R}^*(\gamma_1,\gamma_2,\rho,\sigma,\tau).
\end{equation}
To verify Equation~\eqref{eq:lais:first_reduction} we must show that $(r,\alpha,\kappa) \in \mcf{R}^*(\gamma_1^*,\gamma_2^*,\rho,\sigma,\tau)$ for any $(r,\alpha,\kappa) \in \mcf{R}^*(\gamma_1,\gamma_2,\rho,\sigma,\tau)$. To that goal, observe 
\begin{equation}
\alpha + r < \msf{I}(t\rho,\sigma\tau) - \gamma_2 \leq \msf{I}(t\rho,\sigma\tau) - \gamma_2^*
\end{equation}
and
\begin{equation}
\alpha < \msf{I}(t\rho,\sigma|\tau) - \gamma_1 \leq \msf{I}(t\rho,\sigma\tau) - \gamma_1^*
\end{equation}
for any $(r,\alpha,\kappa) \in \mcf{R}^*(\gamma_1,\gamma_2,\rho,\sigma,\tau)$. Therefore if 
\begin{equation}
\alpha < \min_{\nu } \msf{D}(\nu||q\rho|\sigma\tau) + \left|\kappa_{\nu}^*(\gamma_1^*,\gamma_2^*,\rho,\sigma,\tau) \right|^+
\end{equation}
then $(r,\alpha,\kappa) \in \mcf{R}^*(\gamma_1^*,\gamma_2^*,\rho,\sigma,\tau)$. This though must follow since
\begin{align}
\alpha &< \min_{\nu} \msf{D}( \nu||q\rho|\sigma\tau) + \left|\kappa_{\nu}^*(\gamma_1,\gamma_2,\rho,\sigma,\tau) \right|^+\\
&\leq \msf{D}(\nu^*||q\rho|\sigma\tau) + \left|\kappa_{ \nu^*}^*(\gamma_1,\gamma_2,\rho,\sigma,\tau) \right|^+ \\
&\leq \msf{D}( \nu^*||q\rho|\sigma\tau) + \left|\kappa_{ \nu^*}^*(\gamma_1^*,\gamma_2^*,\rho,\sigma,\tau) \right|^+ 
\end{align}
where
\[
\nu^* = \argmin_{\nu \in \mcf{P}(\mcf{Z}|\mcf{U})} \msf{D}(\nu||q\rho|\sigma\tau) + \abs{\kappa_{\nu}^*(\gamma_1^*,\gamma_2^*,\rho,\sigma,\tau)}^{+}.
\]

\end{IEEEproof}

\subsection{Proof of Theorem \ref{thm:-t2}}\label{app:univcompos}%Thm16

\begin{IEEEproof} 
If $(r,\alpha,\kappa)$ is averagely achievable, then for each $n \in \mathbb{N}$  there exists an
\[
(r_n,\alpha_n,\kappa_n,\epsilon_n,n)
\] 
average authentication code such that
\[
\lim_{n \rightarrow \infty} \left| (r_n,\alpha_n,\kappa_n,\epsilon_n,n) - (r,\alpha,\kappa,0,n)\right|_2 = 0.
\]
Therefore, for any $\beta \in \mathbb{R}$, $r>\beta >0$, there exists a $n' \in \mathbb{N}$ such that $e^{-n^2/4 + 2nr_n} < 1/8$, $r_n > \beta$, and $2^{n\beta} > n^2$ for all $n > n'$, since $r_n$ is converging to a finite $r$.
Thus for some $n' \in \mathbb{N}$ and all $n> n'$ there exists an 
\[
(r_n- \beta, \alpha_n + \beta - 4n^{-1} \logt n, \kappa_n + 2\beta, 2 \epsilon_n,n)  
\]
average authentication code due to the code construction and analysis of Appendix~\ref{app:cc2}.
But if $\epsilon_n \rightarrow 0$, then $2 \epsilon_n \rightarrow 0$.
Hence, this new sequence of codes proves that $(r-\beta,\alpha+ \beta, \kappa + 2\beta)$ is averagely achievable.

\end{IEEEproof}

\subsection{Proof of Theorem \ref{thm:ib}}\label{app:avgregion}%Thm17

By combining Theorem~\ref{thm:lai_strong} and Theorem~\ref{thm:-t2}, $(r,\alpha,\kappa) \in \mcf{C}_{A}$ if there exists a $(r,\alpha,\kappa,r',\alpha',\kappa',\beta) \in \mathbb{R}_+^{7}$ such that 
\begin{align*}
r &= r' - \beta \\
\alpha &= \alpha' + \beta \\
\kappa &= \kappa' + 2\beta \\
\beta &< r' \\
r' + \alpha' &< \msf{I}(t\rho,\sigma\tau) \\
\alpha' &< \min_{\nu \in \mcf{P}(\mcf{Z}|\mcf{U})} \msf{L}(\nu;t\rho,q\rho,\sigma,\tau) \\
\alpha' &< \msf{I}(t\rho,\sigma|\tau)\\
\alpha' - \kappa' &< 0,
\end{align*}
where $\msf{L}^*(\nu; t,q,\sigma,\tau) = \msf{D}(\nu ||q |\sigma\tau) + |\msf{I}(t,\sigma|\tau) - \msf{I}(\nu , \sigma|\tau) + \left| \msf{I}(t\sigma,\tau) - \msf{I}(\nu \sigma,\tau)  \right|^+ |^+  $
for some finite sets $\mcf{U}$, $\mcf{W}$ and distributions $\rho \in \mcf{P}(\mcf{X}|\mcf{U})$, $\sigma \in \mcf{P}(\mcf{U}\gg\mcf{W})$ and $\tau \in \mcf{P}(\mcf{W})$.
The set of triples in Equation~\eqref{eqn:avgregion} are obtained by applying Fourier-Motzkin elimination to remove $r',\alpha',\kappa'$ and $\beta$ to this preceding set of equations, and taking the closure.

\subsection{Code construction and analysis for Theorem~\ref{thm:lai_strong}}\label{app:cc1} %Thm 15
Given finite sets $\mcf{U}$ and $\mcf{W}$, distributions $\rho \in \mcf{P}(\mcf{X}|\mcf{U})$, $\sigma \in \mcf{P}_{n} (\mcf{U} \gg \mcf{W})$, and $\tau \in \mcf{P}_n(\mcf{W})$, constant $\gamma > 0$, and large enough $n$, the following code construction results in a $(r,\alpha ,\kappa, \epsilon,n)$-AA code for DM-ASC$(t,q,1)$, where 
\[
\epsilon \leq \max_{\mu \in \mcf{P}(\mcf{Y}|\mcf{U})} 2^{- n\msf{D}(\mu||t\rho|\sigma\tau) + O(\logt n)} \max \left(  2^{n\left| \tilde r + \tilde \kappa -  \msf{I}(\mu,\sigma|\tau)\right|^-  } , 2^{n \left| \tilde r + \hat r + \tilde \kappa - \msf{I}(\mu,\sigma\tau)\right|^- }  \right),
\]
for all $r \leq \tilde r + \hat r$, $\kappa \geq \tilde \kappa$, and
\[
\alpha \geq \min_{\nu \in \mcf{P}(\mcf{Z}|\mcf{U})} \msf{D}(\nu||q\rho|\sigma\tau) + \left| \tilde \kappa  - \left| \msf{I}(\nu,\sigma|\tau) - \tilde r -  \left|\hat r - \msf{I}(\nu\sigma,\tau) \right|^+ \right|^+ \right|^+ - O(n^{-1}\logt n),
\]
with the order terms dependent only on the cardinalities of $\mcf{U},~\mcf{W},~\mcf{X},~\mcf{Y}$ and $\mcf{Z}$

\subsubsection{Code Construction}\label{app:cc1:cc}~\\
The code construction involves rate splitting. That is the message set $\mcf{M} \defn \set{1, \dots, 2^{nr}}$ is represented by the product of sets $\mcf{\tilde M} \defn \set{1, \dots, 2^{n\tilde r}}$ and $\mcf{\hat M} \defn \set{1, \dots, 2^{n\hat r}}$, such that $\tilde r + \hat r  = r$. In other words $\mcf{M} = \mcf{\hat M} \times \mcf{\tilde M}$. The set of keys is denoted $\mcf{K} = \set{1, \dots, 2^{n\tilde \kappa}}$. 

Let $\mcf{U}$ and $\mcf{W}$ be finite and fix $\rho \in \mcf{P}(\mcf{X}|\mcf{U})$, $\sigma \in \mcf{P}_n(\mcf{U}\gg\mcf{W})$, and $\tau \in \mcf{P}_n(\mcf{W})$.

\noindent \textbf{Key Establishment:} Before communication, Alice and Bob jointly select a key, $k$, uniformly at random from $\mcf{K}$.

\noindent  \textbf{Random codebook generation:} Independently for each $\hat m \in \mcf{\hat M}$, Alice selects a codeword $w^n(\hat m)$ uniformly at random from $\mcf{T}_{\tau}^n$. Then, independently for each $(\hat m,\tilde m, k)  \in \mcf{\hat M}\times \mcf{\tilde M} \times \mcf{K}$, Alice independently selects a codeword $u^n(\hat m, \tilde m, k)$ uniformly at random from $\mcf{T}^n_{\sigma}(w^n(\hat m))$.

\noindent  \textbf{Encoder:} Given Alice observes $\hat m, \tilde m$, and has key $k$, Alice selects codeword $x^n$ to transmit according to 
\begin{equation}\label{eq:encoder}
f(x^n|\hat m,\tilde m,k) \defn \rho^{\otimes n}(x^n|u^n(\hat m, \tilde m, k) ).
\end{equation}
Note then that $Y^n|\hat M,\tilde M, K$ and $Z^n|\hat M, \tilde M, K$ will be distributed $(t\rho)^{\otimes n}(y^n|u^n(\hat m,\tilde m,k))$ and $(q\rho)^{\otimes n}(z^n|u^n(\hat m,\tilde m,k))$ respectively.

\noindent  \textbf{Decoder:} Given Bob observes $y^n$ and has key $k$, Bob chooses $\hat M', \tilde M'$ according to the distribution
\begin{equation}\label{eq:decoder1}
\varphi(\hat m', \tilde m'|y^n,k) \defn \begin{cases}
1 & \text{ if } (t\rho)^{\otimes n} ( y^n|u^n(\hat m',\tilde m', k)) > (t\rho)^{\otimes n} ( y^n|u^n(i, j,l))  \\
& ~~ \forall (i,j,l)  \in \mcf{\hat M}\times \mcf{\tilde M} \times \mcf{K} - \set{(\hat m',\tilde m', k)}\\
0 & \text{ otherwise}
\end{cases},
\end{equation}
for all $(\hat m,\tilde m, k,y^n) \in \mcf{\hat M} \times \mcf{\tilde M} \times \mcf{K} \times \mcf{Y}^n$, and 
\begin{equation}\label{eq:decoder2}
\varphi(\mbf{!}|y^n,k) \defn \begin{cases}
1 & \text{ if } \sum_{(\hat m,\tilde m) \in \mcf{\hat M} \times \mcf{\tilde M}} \varphi(\hat m',\tilde m'|y^n,k) = 0\\
0 & \text{ otherwise}
\end{cases},
\end{equation}
for all $(y^n, k) \in \mcf{Y}^n \times \mcf{K}$. That is for a given $y^n$, Bob declares $(\hat m',\tilde m')$ to be the message sent if $ (t\rho)^{\otimes n}(y^n|u^n(\hat m', \tilde m', k) > (t\rho)^{\otimes n} ( y^n|u^n(i, j, k'))$ for all $(i,j,k') \in \mcf{\hat M}\times \mcf{\tilde M} \times \mcf{K} - \set{(\hat m',\tilde m',k)}$. If this maximum is not unique, or if the maximum occurs for $k' \neq k$, then $\mbf{!}$ is declared.

\subsubsection{Message error analysis}\label{app:cc1:mea}~\\
Let $F,\Phi$ be the random variables that represent the randomly chosen encoder and decoder. We will show
\begin{equation} \label{eq:proof_lai:fin2}
E_{F,\Phi}(\varepsilon_{f,\varphi}) \leq \max_{\mu \in \mcf{P}(\mcf{Y}|\mcf{U})} 2^{- n\msf{D}(\mu||t\rho|\sigma\tau) + O(\logt n)} \max \left(  2^{n\left| \tilde r + \tilde \kappa -  \msf{I}(\mu,\sigma|\tau)\right|^-  } , 2^{n \left| \tilde r + \hat r + \tilde \kappa - \msf{I}(\mu,\sigma\tau)\right|^- }  \right).
\end{equation}
In turn then, 
%with probability $1-n^{-1}$ a code is selected such that $\varepsilon_{f,\varphi} < n E_{F,$ is less than $n$ times the right hand side of Equation~\eqref{eq:proof_lai:fin2}. In other words
\begin{equation} \label{eq:proof_lai:fin2alt}
p_{F,\Phi} \left( \set{f, \varphi:  \varepsilon_{f,\varphi} \geq n E_{F,\Phi}(\varepsilon_{f,\varphi})  } \right) < n^{-1}
\end{equation}
follows directly from Markov's inequality. 

To aid in proving Equation~\eqref{eq:proof_lai:fin2}, define sets 
\[
\mcf{D}(\mu) \defn \set{ \mu' : \msf{H}(\mu'|\sigma\tau) + \msf{D}(\mu'||t\rho|\sigma\tau) \leq \msf{H}(\mu|\sigma\tau) + \msf{D}(\mu||t\rho|\sigma\tau) },
\]
and 
\[
\mcf{E}(u^n,y^n) \defn \set{\hat u^n \in \mcf{T}^n_{\sigma\tau} : y^n \in \cup_{\mu' \in \mcf{D}(\mu^*)} \mcf{T}^n_{\mu'}(\hat u^n)   },
\]
where $\mu^* \defn p_{y^n|u^n}$. With these definitions in hand, observe
\begin{equation} \label{eq:proof_lai:error2}
\varepsilon_{f,\varphi} = \hspace{-15pt} \sum_{\hat m,\tilde m, k,u^n,y^n} 2^{-n( \hat r + \tilde r + \kappa)} (t\rho)^{\otimes n}(y^n|u^n) \min \left( \sum_{(i,j,l) \neq (\hat m,\tilde m,k)} \idc{u^n = u^n(\hat m, \tilde m, k) } \idc{u^n(i,j,l) \in \mcf{E}(u^n,y^n) } , 1 \right).
\end{equation} 
Indeed, if $u^n(\hat m, \tilde m , k) = u^n$, and there exists a $u^n(i,j,l) \in \mcf{E}(u^n,y^n)$ for some $(i,j,l) \neq (\hat m,\tilde m, k)$, then an error will occur since
\[
(t\rho)^{\otimes n}(y^n| u^n(i,j,l)) = 2^{-n(\msf{H}(\mu'|\sigma\tau) + \msf{D}(\mu' || t\rho | \sigma\tau)) } \geq 2^{-n(\msf{H}(\mu|\sigma\tau) + \msf{D}(\mu || t\rho | \sigma\tau)) } = (t\rho)^{\otimes n}(y^n|u^n(\hat m,\tilde m, k))
\]
where $\mu'\defn p_{y^n|u^n(i,j,l)}$ and $\mu \defn p_{y^n|u^n}$. Splitting the summation of $y^n$ depending on the empirical distribution $y^n|u^n$ leads to 
\begin{align} 
\varepsilon_{f,\varphi} &=\sum_{\hat m,\tilde m, k,u^n}\sum_{\mu \in \mcf{P}_n(\mcf{Y}|\mcf{U})} \sum_{y^n \in \mcf{T}^n_{\mu}(u^n)} 2^{-n( \hat r  + \tilde r+ \tilde \kappa + \msf{H}(\mu|\sigma\tau) + \msf{D}(\mu||t\rho|\sigma\tau) )} \notag \\
&\hspace{10pt} \times \min\left( \sum_{(i,j,l) \neq (\hat m,\tilde m,k)} \idc{u^n = u^n(\hat m, \tilde m, k) } \idc{u^n(i,j,l) \in \mcf{E}(u^n,y^n) },1 \right).\label{eq:proof_lai:error3}
\end{align}
It is important to note that the value of $\mu$ fixes $p_{y^n}$, since $p_{y^n|w^n} = \mu \sigma$ by Lemma~\ref{lem:unique}, and thus $p_{y^n} = p_{y^n|w^n}p_{w^n} = \mu\sigma\tau$.

Having found a formula for $\varepsilon_{f,\varphi}$, that $E_{F,\Phi} [\varepsilon_{f,\varphi}]$ is less than 
\begin{align} 
& \sum_{\mu \in \mcf{P}_n(\mcf{Y}|\mcf{U})}  \sum_{\hat m, \tilde m,k,u^n}\sum_{y^n \in \mcf{T}^n_{\mu}(u^n)} 2^{-n( \hat r + \tilde r+ \tilde \kappa + \msf{H}(\mu|\sigma\tau) + \msf{D}(\mu||t\rho|\sigma\tau) )} \notag \\
&\hspace{10pt} \times\min \left( \sum_{(i,j,l) \neq (\hat m,\tilde m, k)} E_{F,\Phi} \left[ \idc{u^n = u^n(\hat m, \tilde m, k) } \idc{u^n(i,j,l) \in \mcf{E}(u^n,y^n)}  \right] ,1 \right), \label{eq:proof_lai:avg}
\end{align}
follows by applying the expectation operator to~\eqref{eq:proof_lai:error3}, and then making use of the linearity of the expected value, and that the expected value of the minimum is less than or equal to the minimums of the expected values. But note that 
\begin{equation}\label{eq:cc1:mea:1}
E_{F,\Phi} \left[ \idc{u^n = u^n(\hat m, \tilde m, k) } \idc{u^n(i,j,l) \in \mcf{E}(u^n,y^n)}  \right] = \Pr\left( U^n(\hat m, \tilde m, k) = u^n, U^n(i,j,l) \in \mcf{E}(u^n,y^n) \right),
\end{equation}
where $U^n(\hat m ,\tilde m, k)$ is the random variable denoting the $u^n$ chosen for $(\hat m,\tilde m, k)$. If $\hat m \neq i$, then $U^n(\hat m,\tilde m, k)$ and $U^n(i,j,l)$ are independently chosen over $\mcf{T}_{\sigma\tau}^n$, and thus in this case Equation~\eqref{eq:cc1:mea:1} equals
\begin{align}
&\Pr\left( U^n(\hat m, \tilde m, k) = u^n \right) \Pr \left( U^n(i,j,l) \in \mcf{E}(u^n,y^n) \right) \notag\\
&=\abs{\mcf{T}_{\sigma\tau}^n}^{-1} \sum_{\mu' \in \mcf{D}(\mu)} \Pr( y^n \in \mcf{T}_{\mu'}^n(U^n) ) \notag\\
&\leq 2^{-n \msf{H}(\sigma\tau) + O(\logt n) } \sum_{\mu' \in \mcf{D}(\mu): \mu'\sigma\tau = \mu\sigma\tau } 2^{-n \msf{I}(\mu',\sigma\tau)}\label{eq:cc1:mea:1.1},
\end{align}
by Lemma~\ref{lem:obvious}. Note the restriction in Equation~\eqref{eq:cc1:mea:1.1} is due to the value of $\mu$ fixing the value of $\mu\sigma\tau$. Otherwise if $\hat m = i$, then $U^n(\hat m, \tilde m, k)$ and $U^n(i,j,l)$ must both belong to $\mcf{T}^n_{\sigma}(w^n(\hat m))$. Thus when $\hat m = i$, Equation~\eqref{eq:cc1:mea:1} equals
\begin{align}
&\Pr\left( U^n(\hat m, \tilde m, k) = u^n \right) \Pr \left( U^n(i,j,l) \in \mcf{E}(u^n,y^n) \middle| W^n = w^n \right) \notag\\
&=\abs{\mcf{T}_{\sigma\tau}^n}^{-1} \sum_{\mu' \in \mcf{D}(\mu): \mu'\sigma = \mu\sigma} \Pr( y^n \in \mcf{T}_{\mu'}^n(U^n) | U^n(i,j,l) \in \mcf{T}^n_{\sigma}(w^n(\hat m)) ) \notag\\
&\leq  2^{-n \msf{H}(\sigma\tau) + O(\logt n) } \sum_{\mu' \in \mcf{D}(\mu):\mu'\sigma = \mu\sigma} 2^{-n \msf{I}(\mu',\sigma|\tau)}\label{eq:cc1:mea:1.2}
\end{align}
by Lemma~\ref{lem:obvious}. As before, the restriction of $\mu$ is because $\mu$ fixes the value of $p_{y^n|w^n}$.

Hence we have $E_{F,\Phi}[\varepsilon_{f,\varphi}] $ is upper bounded by
\begin{equation} \label{eq:proof_lai:avg2}
\sum_{\mu \in \mcf{P}_n(\mcf{Y}|\mcf{U})}  \hspace{-10pt}  2^{-n\msf{D}(\mu||t\rho|\sigma\tau)  + O(\logt n) } \left(\sum_{\mu' \in \mcf{D}(\mu) : \mu'\sigma = \mu \sigma} \hspace{-20pt} 2^{n\left| \tilde r +\tilde \kappa -  \msf{I}(\mu',\sigma|\tau) \right|^- } + \hspace{-20pt}\sum_{\mu' \in \mcf{D}(\mu) : \mu'\sigma\tau = \mu \sigma \tau} \hspace{-20pt} 2^{n \left| \hat r + \tilde r + \tilde \kappa - \msf{I}(\mu',\sigma\tau)\right|^-} \right) ,
\end{equation}
from substituting Equations~\eqref{eq:cc1:mea:1.1} and~\eqref{eq:cc1:mea:1.2} into Equation~\eqref{eq:proof_lai:avg}, and then carrying out the summations. To conclude the proof we will show that the maximum summand in Equation~\eqref{eq:proof_lai:avg2} occurs when $\mu' = \mu$. Because of this
\begin{equation} \label{eq:proof_lai:fin}
E_{F,\Phi}(\epsilon_{f,\varphi}) \leq \max_{\mu \in \mcf{P}_n(\mcf{Y}|\mcf{U})} 2^{- n\msf{D}(\mu||t\rho|\sigma\tau) + O(\logt n)} \max \left(  2^{n\left| \tilde r +\tilde  \kappa -  \msf{I}(\mu,\sigma|\tau)\right|^-  } , 2^{n \left| \tilde r + \hat r + \tilde \kappa - \msf{I}(\mu,\sigma\tau)\right|^- }  \right),
\end{equation}
since the number of summation terms grows at most polynomial with $n$. Equation~\eqref{eq:proof_lai:fin2} (which is what we set out to prove) then follows since $\mcf{P}_{n}(\mcf{Y}|\mcf{U}) \subset \mcf{P}(\mcf{Y}|\mcf{U})$ for all $n \in \mathbb{N} $.

Thus we conclude with demonstrating the maximum summand in Equation~\eqref{eq:proof_lai:avg2} to occur when $\mu = \mu'$. To verify observe that 
\[
\msf{D}(\mu'||t\rho|\sigma\tau) \leq \msf{H}(\mu|\sigma\tau) - \msf{H}(\mu'|\sigma\tau) + \msf{D}(\mu||t\rho|\sigma\tau)
\]
for all $\mu' \in \mcf{D}(\mu)$, and thus if $\msf{H}(\mu'|\sigma\tau)$ is greater than $\msf{H}(\mu|\sigma\tau)$, then $\msf{D}(\mu'||t\rho|\sigma\tau)$ is less than $\msf{D}(\mu||t\rho|\sigma\tau)$. From this our assumption is proved since 
\begin{equation}\label{eq:proof_lai:verify}
\msf{H}(\mu'|\sigma\tau) - \msf{D}(\mu||t\rho|\sigma\tau) \leq \max \left(  \msf{H}(\mu|\sigma\tau) - \msf{D}(\mu||t\rho|\sigma\tau) , \msf{H}(\mu'|\sigma\tau) - \msf{D}(\mu'||t\rho|\sigma\tau) \right),
\end{equation}
and either $\msf{H}(\mu'\sigma|\tau) = \msf{H}(\mu\sigma|\tau)$ or $\msf{H}(\mu'\sigma\tau) = \msf{H}(\mu\sigma\tau)$ for the summands.

\subsubsection{Type I error analysis}\label{app:cc1:t1e}~\\
Before presenting the sufficient conditions, it is important to observe that the type I error can be simplified to 
\begin{equation}\label{eq:avg:t1e_redux}
\omega_{f,\varphi} = \sum_{z^n \in \mcf{Z}^n} \max_{k} p(z^n,k).
\end{equation}
This follows by directly inserting the optimal $\psi \in \mcf{P}(\mcf{Y}^n|\mcf{Z}^n)$ into the definition of the average type I error. A global maximum occurs when $\psi$ takes the form $\psi(y^n|z^n) = \idc{y^n = g(z^n)}$, where $g: \mcf{Z}^n \rightarrow \mcf{Y}^n$ is any function for which 
\[
\sum_{k } p(z^n,k) \varphi(\mcf{M}|g(z^n),k)  = \max_{y^n \in \mcf{Y}^n}\sum_{k } p(z^n,k) \varphi(\mcf{M}|y^n,k) ,
\]
for all $z^n \in \mcf{Z}^n$.
%This follows from re-writing Equation~\eqref{eq:avg:t1e} as
%\begin{equation}
%\sum_{z^n} \max_{\psi} \sum_{y^n} \psi(y^n|z^n) \sum_{k} p_{Z^n,K}(z^n,k) \varphi(\mcf{M}|y^n,k)
%\end{equation}
Thus Equation~\eqref{eq:avg:t1e_redux} is valid since $\varphi(\mcf{M}|y^n,k) = 1$ for at most one $k$ by the code construction.

As will be shown later, if
\begin{equation}\label{eq:avg:suff:1}
\logt \abs{\set{(\hat m,\tilde m,k): z^n \in \mcf{T}^n_{\nu}(u^n(\hat m,\tilde m, k) ) }} \dot= n \left| \left| \hat r - \msf{I}(\nu\sigma,\tau) \right|^+ + \tilde r + \tilde \kappa - \msf{I}(\nu,\sigma|\tau) \right|^+ + O(\logt n)
\end{equation}
and
\begin{equation}\label{eq:avg:suff:2}
\logt \max_k \abs{\set{(\hat m,\tilde m): z^n \in \mcf{T}^n_{\nu}(u^n(\hat m,\tilde m, k) ) }} \dot= n \left| \left| \hat r - \msf{I}(\nu\sigma,\tau) \right|^+ + \tilde r - \msf{I}(\nu,\sigma|\tau) \right|^+ + O(\logt n)
\end{equation}
for all $\nu\in \mcf{P}_{n} (\mcf{Z}|\mcf{U})$ and $z^n \in \cup_{\hat m,\tilde m, k} \mcf{T}^n_{\nu}(u^n(\hat m,\tilde m, k))$,
then
\begin{equation}\label{eq:avg:t1e:ub}
\omega_{f,\varphi} < \max_{\nu \in \mcf{P}_{n}(\mcf{Z}|\mcf{U}) } 2^{-n \left( D(\nu||q\rho|\sigma\tau) + \left| \tilde \kappa - \left| I(\nu,\sigma|\tau) - \tilde r - \left| \hat r - \msf{I}(\nu\sigma,\tau) \right|^+\right|^+ \right|^+ \right) + O(\logt n)}.
\end{equation} 
We now show that $1- e^{-n^2/4 + O(\logt n)}$ lower bounds the probability of randomly selecting a code that satisfies Conditions~\eqref{eq:avg:suff:1} and~\eqref{eq:avg:suff:2}. First observe that $1-2\abs{\mcf{Z}}^ne^{-n^2/4} = 1-e^{-n^2/4 + O(n)}$ lower bounds the probability that a code is chosen for which
\begin{equation} \label{eq:avg:suff:1.1}
\abs{\set{\hat m : z^n \in \mcf{T}^n_{\nu\sigma}(w^n(\hat m)) }} \dot = 2^{n \left| \hat r - \msf{I}(\nu\sigma,\tau) \right|^+ + O(\logt n) }
\end{equation}
for all $z^n \in \bigcup_{\hat m} \mcf{T}^n_{\nu\sigma}(w^n(\hat m))$. This observation follows because for each $\hat m$, $w^n(\hat m)$ is independently chosen according to a uniform distribution over $\mcf{T}_{\tau}^n$. Hence 
\[
\Pr \left( z^n \in \mcf{T}^n_{\nu\sigma}(W^n(\hat m)) \right) \dot=2^{-n\msf{I}(\nu\sigma,\tau) + O(\logt n)}
\]
for all $z^n \in \mcf{T}_{\nu\sigma\tau}^n$  by Lemma~\ref{lem:obvious}, and therefore $1-2e^{-n^2/4}$ lower bounds the probability of selecting a code that satisfies Equation~\eqref{eq:avg:suff:1.1} for any $z^n \in \mcf{T}_{\nu\sigma\tau}^n$ by Corollary~\ref{cor:ck}. Next, if the values of $w^n(\hat m)$ are chosen such that Equation~\eqref{eq:avg:suff:1.1} is true for all $z^n \in \cup_{\hat m \in \mcf{\hat M}} \mcf{T}_{\nu\sigma}^n(w^n(\hat m))$, then $1-4 \abs{\mcf{K}} \abs{\mcf{Z}}^n e^{-n^2/4}$ lower bounds the probability of a code being selected that also satisfies Conditions~\eqref{eq:avg:suff:1} and~\eqref{eq:avg:suff:2}. Indeed for each $z^n \in \cup_{\hat m \in \mcf{\hat M}} \mcf{T}_{\nu\sigma}^n(w^n(\hat m))$, there will be $2^{n \left( \tilde r + \tilde \kappa + \left| \hat r - \msf{I}(\nu\sigma,\tau) \right|^+ \right) + O(\logt n) }$ values $(\hat m,\tilde m,k)$ for which $u^n(\hat m, \tilde m, k)$ will independently be chosen uniformly at random from a $\mcf{T}_{\nu\sigma}^n(w^n(\hat m))$ such that  $z^n \in \mcf{T}_{\nu\sigma}^n(w^n(\hat m))$. Likewise for each $z^n \in \cup_{\hat m \in \mcf{\hat M}} \mcf{T}_{\nu\sigma}^n(w^n(\hat m))$ and $k \in \mcf{K}$, there will be $2^{n \left( \tilde r  + \left| \hat r - \msf{I}(\nu\sigma,\tau) \right|^+ \right) + O(\logt n) }$ values $(\hat m,\tilde m,k)$ for which $u^n(\hat m, \tilde m, k)$ will independently be chosen uniformly at random from a $\mcf{T}_{\sigma}^n(w^n(\hat m))$ such that  $z^n \in \mcf{T}_{\nu\sigma}^n(w^n(\hat m))$. Furthermore, for each of these randomly selected values
\[
\Pr \left( z^n \in \mcf{T}^n_{\nu}(U^n(\hat m,\tilde m, k)) \middle| W^n(\hat m) = w^n \right) \dot=2^{-n\msf{I}(\nu,\sigma|\tau) + O(\logt n)}
\]
if $z^n \in \mcf{T}_{\nu\sigma}^n(w^n)$ by Lemma~\ref{lem:obvious}, and $0$ otherwise due to Lemma~\ref{lem:unique}.  Thus with probability greater than $1-2\abs{\mcf{Z}}^ne^{-n^2/4}$, the values of $u^n(\hat m, \tilde m, k)$ are chosen that satisfy Condition~\eqref{eq:avg:suff:1} for all $z^n \in \cup_{(\hat m,\tilde m, k)} \mcf{T}_{\nu}(u^n(\hat m,\tilde m, k))$ by Corollary~\ref{cor:ck}. Likewise with probability greater than $1-2\abs{\mcf{Z}}^n \abs{\mcf{K}} e^{-n^2/4}$ the values of $u^n(\hat m, \tilde m, k)$ are chosen that satisfy Condition~\eqref{eq:avg:suff:1} for all $z^n \in \cup_{(\hat m,\tilde m, k)} \mcf{T}_{\nu}(u^n(\hat m,\tilde m, k))$ by Corollary~\ref{cor:ck}. Thus, $1-e^{-n^2/4 + O(n)}$ must lower bound the probability of choosing a code that satisfies Conditions~\eqref{eq:avg:suff:1} and~\eqref{eq:avg:suff:2} for all $\nu \in \mcf{P}_{n}(\mcf{Z}|\mcf{U})$ and $z^n \in \cup_{\hat m, \tilde m, k} \mcf{T}_{\nu}^n(u^n(\hat m, \tilde m, k))$.

We now proceed with showing conditions~\eqref{eq:avg:suff:1} and~\eqref{eq:avg:suff:2} cause~\eqref{eq:avg:t1e:ub}. Set \[\mcf{Z}^n_{f}(\nu)\defn \set{ z^n \in \cup_{\hat m, \tilde m, k} \mcf{T}^n_{\nu}(u^n(\hat m, \tilde m, k))}.\] The type I error for $f,\varphi$ can be bounded as follows:
\begin{align}
\omega_{f,\varphi} &= \sum_{z^n}  2^{-n (\hat r + \tilde r + \tilde \kappa) }  \max_{k} \sum_{\hat m,\tilde m, u^n} (q\rho)^{\otimes n}(z^n|u^n) \idc{u^n = u^n(\hat m,\tilde m, k)} \label{eq:avg:t1e:3}\\
&\leq \sum_{z^n}  2^{-n (\hat r+ \tilde r + \tilde \kappa) }  \max_{k} \sum_{\nu \in \mcf{P}_{n}(\mcf{Z}|\mcf{U}) } \abs{\set{(\hat m,\tilde m): z^n \in \mcf{T}^n_{\nu}(u^n(\hat m,\tilde m, k) ) }} 2^{-n(\msf{H}(\nu|\sigma\tau) + \msf{D}(\nu||q\rho|\sigma\tau))}  \label{eq:avg:t1e:4}\\
&\leq  2^{-n (\hat r+ \tilde r + \tilde \kappa) } \sum_{\nu \in \mcf{P}_{n}(\mcf{Z}|\mcf{U}) }  \sum_{z^n\in \mcf{Z}^n_f(\nu) }  \max_{k}  \abs{\set{(\hat m,\tilde m): z^n \in \mcf{T}^n_{\nu}(u^n(\hat m,\tilde m, k) ) }} 2^{-n(\msf{H}(\nu|\sigma\tau) + \msf{D}(\nu||q\rho|\sigma\tau))}
\label{eq:avg:t1e:5}\\
&\leq 2^{-n (\hat r + \tilde r + \tilde \kappa)} \sum_{\nu \in \mcf{P}_{n}(\mcf{Z}|\mcf{U}) } \sum_{z^n \in \mcf{Z}^n_f(\nu)}  2^{n \left| \tilde r - \msf{I}(\nu,\sigma|\tau) + \left| \hat r - \msf{I}(\nu\sigma,\tau) \right|^+ \right|^+ + O(\logt n)} 2^{-n(\msf{H}(\nu|\sigma\tau) + \msf{D}(\nu||q\rho|\sigma\tau))} \label{eq:avg:t1e:5.1}\\
&\leq 2^{-n (\hat r + \tilde r + \tilde \kappa)} \max_{\nu \in \mcf{P}_{n}(\mcf{Z}|\mcf{U}) } \abs{\mcf{Z}^n_f(\nu)}  2^{n \left| \tilde r - \msf{I}(\nu,\sigma|\tau) + \left| \hat r - \msf{I}(\nu\sigma,\tau) \right|^+ \right|^+ + O(\logt n)} 2^{-n(\msf{H}(\nu|\sigma\tau) + \msf{D}(\nu||q\rho|\sigma\tau))} \label{eq:avg:t1e:5.2}
\end{align}
where~\eqref{eq:avg:t1e:3} follows from Equation~\eqref{eq:avg:t1e_redux} because
\[
p(z^n,k) = \sum_{\hat m, \tilde m} p(z^n|\hat m, \tilde m , k) 2^{-n (\hat r + \tilde r + \tilde \kappa)} = \sum_{\hat m, \tilde m,u^n} (q\rho)^{\otimes n}(z^n|u^n) \idc{u^n = u^n(\hat m, \tilde m , k)} 2^{-n (\hat r + \tilde r + \tilde \kappa)},
\]
\eqref{eq:avg:t1e:4} follows by further sub-dividing the summation based upon $p_{z^n|u^n}$, and then performing the sum over all $(\hat m,\tilde m,u^n)$ for a fixed $p_{z^n|u^n}$,~\eqref{eq:avg:t1e:5} is from moving the maximum inside the summation (which does not decrease the value) and recognizing $\abs{\set{(\hat m,\tilde m): z^n \in \mcf{T}^n_{\nu}(u^n(\hat m,\tilde m, k) ) }}$ is only non-zero if $z^n \in \mcf{Z}_{f}^n(\nu)$,~\eqref{eq:avg:t1e:5.1} is by Condition~\eqref{eq:avg:suff:2}, and finally~\eqref{eq:avg:t1e:5.2} results from performing the summation over $\mcf{Z}_f^n(\nu)$ and because $
\sum_{\nu \in \mcf{P}_{n}(\mcf{Z}|\mcf{U})} \cdot \leq \abs{\mcf{P}_{n}(\mcf{Z}|\mcf{U})} \max_{\nu \in \mcf{P}_{n}(\mcf{Z}|\mcf{U}) } \cdot .$

To finish the proof, observe
\begin{align*}
\abs{\mcf{Z}^n_f(\nu)} &\leq \frac{2^{n(\hat r +\tilde r+ \kappa)}2^{n\msf{H}(\nu|\sigma\tau) + O(\logt n) } }{\min_{z^n \in \mcf{Z}^n_f(\nu)} \abs{\set{\hat m, \tilde m , k : z^n \in \mcf{T}^n_{\nu}(u^n(\hat m, \tilde m , k)) }}}
\end{align*}
by Lemma~\ref{lem:tbcut}. Hence
\begin{equation}\label{eq:avg:t1e:|Z|}
\abs{\mcf{Z}^n_f(\nu)} \leq  2^{O(\logt n)} \frac{2^{n(\hat r + \tilde r + \tilde \kappa)}2^{n\msf{H}(\nu|\sigma\tau)  } }{2^{n\left| \tilde r + \tilde \kappa - \msf{I}(\nu,\sigma|\tau) + \left| \hat r - \msf{I}(\nu\sigma,\tau) \right|^+ \right|^+  }}
\end{equation}
by Condition~\eqref{eq:avg:suff:1}. Equation~\eqref{eq:avg:t1e:ub} follows by combining Equations~\eqref{eq:avg:t1e:5.2} and~\eqref{eq:avg:t1e:|Z|}.

\subsection{Code construction and analysis for Theorem~\ref{thm:-t2}}\label{sec:-t}\label{app:cc2} %Thm 16

Let $(f,\varphi)$ be a code with rate $r$, block-length $n$, key requirement $\kappa$, and assume $r > \beta > 2 n^{-1}\logt n$.
The following construction and analysis guarantees the existence of a code $(f',\varphi')$ if 
\[
e^{-n^2/4 + 2nr } < 1/8.
\]
In specific, if $(f,\varphi)$ is a $(r,\alpha,\kappa,\epsilon,n)$-AA code, then $(f',\varphi')$ will be a 
\[
(r-\beta, \alpha + \beta - 2n^{-1} \logt ne^3 , \kappa + 2 \beta, 2 \epsilon,n)
\]
AA code.

\subsubsection{Code Construction}\label{app:cc2:cc}~\\
\noindent \textbf{Original code:} Let $(f,\varphi) \in \mcf{P}(\mcf{X}^n|\mcf{M}, \mcf{K}_1) \times \mcf{P}(\mcf{M} \cup \set{\mbf{!}} | \mcf{Y}^n, \mcf{K}_1)$ be a code provided to Alice and Bob, where $\mcf{M} \defn \set{1,\dots,2^{nr}}$ and $\mcf{K}_1 \defn \set{1 ,\dots, 2^{n \kappa}}$. 

\noindent \textbf{Key Establishment:} Before communication, Alice and Bob jointly select a key, $(k_1,k_2)$, uniformly at random from $\mcf{K}_1\times \mcf{K}_2$, where $\mcf{K}_2 \defn \set{1,\dots,2^{2n\beta}}$. 

\noindent  \textbf{Random codebook generation:} Let $\mcf{M}' \defn \set{1 , \dots, 2^{n (r-\beta)}}$. Alice independently for every $k_2 \in \mcf{K}_2$ selects a set $\mcf{M}(k_2)$ uniformly at random from the set of all $2^{n( r - \beta)}$-element subsets of $\mcf{M}$. Next, Alice independently for every $k_2 \in \mcf{K}_2$ chooses an invertible mapping $g_{k_2}: \mcf{M}' \rightarrow \mcf{M}(k_2)$ uniformly at random from the set of invertible mappings on $\mcf{M}' \rightarrow \mcf{M}(k_2)$. Her choices are publicly shared.

\noindent  \textbf{Encoder:}
\[
f'(x^n | m',k_1,k_2) \defn f(x^n | g_{k_2}(m'),k_1)
\]
for all $(x^n,m',k_1,k_2) \in \mcf{X}^n \times \mcf{M}' \times \mcf{K}_1 \times \mcf{K}_2$. 

\noindent  \textbf{Decoder:}
\[
\varphi'(m'|y^n,k_1,k_2) = \begin{cases}
\varphi(g_{k_2}(m')|y^n,k_1) &\text{ if } m' \neq \mbf{!}  \\
\varphi(\mbf{!} | y^n,k) + \varphi(\mcf{M} - \mcf{M}(k_2)|y^n,k_1) &\text{ otherwise} 
\end{cases},
\]
for all $(y^n,k_1, k_2) \in \mcf{Y}^n\times \mcf{K}_1 \times \mcf{K}_2$ and  $m' \in \mcf{M}' \cup \mbf{!}$.

\subsubsection{Message error analysis}\label{app:cc2:mea}~\\
The average probability of message error, when averaged over all possible $(f',\varphi')$, is equal to the probability of message error for $(f,\varphi)$. Indeed, this is because the probability of message error for $(f',\varphi')$ and a given $m',k_1,k_2$ is equal to the probability of message error of $(f,\varphi)$ for some $ m,k_1$ since
\begin{align}
\varepsilon_{f',\varphi'}(m',k_1,k_2) &= 1- \sum_{y^n,x^n} \varphi'(m'|y^n,k_1,k_2)  t^n(y^n|x^n) f'(x^n|m',k_1,k_2) \notag \\
&= 1- \sum_{y^n,x^n} \varphi(g_{k_2}(m')|y^n,k_1)  t^n(y^n|x^n) f(x^n|g_{k_2}(m'),k_1) \notag \\
&= \varepsilon_{f,\varphi}(g_{k_2}(m'),k_1).
\end{align}
Therefore
\begin{equation}
E_{F',\Phi'}[ \varepsilon_{f',\varphi'} ]  = \sum_{m' \in \mcf{M}', k_1 \in \mcf{K}_1 } 2^{-n(r - \beta + \kappa)} \left( \sum_{m \in \mcf{M}} 2^{-n r } \varepsilon_{f,\varphi}(m,k_1)\right) = \varepsilon_{f,\varphi} ,
\end{equation}
since, for a given $k_1$, the value of $m$ that corresponds to $m'$ is uniformly chosen over all $\mcf{M} \times \mcf{K}_1$. Thus $\varepsilon_{f',\varphi'} < 2 \varepsilon_{f,\varphi}$ for at least half of the possible codes due to Markov's inequality.

\subsubsection{Type I error analysis}\label{app:cc2:t1e}~\\
In order to guarantee
\begin{equation}\label{eq:thm:-t:tt1e}
\omega_{f',\varphi'} <  \omega_{f,\varphi}2^{-n \beta  + 2\logt ne^3 },
\end{equation}
it is sufficient to show
\begin{equation}\label{eq:thm:-t:suff1}
e^{-2} 2^{n \beta} < \abs{\set{k_2 :m \in \mcf{M}(k_2)}} < e^2 2^{n \beta}
\end{equation}
and
\begin{equation}\label{eq:thm:-t:suff2}
\abs{\set{k_2 :(m,\tilde m) \in \mcf{M}(k_2) \times \mcf{M}(k_2)} } \leq (ne)^2,
\end{equation}
for all $(m,\tilde m) \in \mcf{M} \times \mcf{M}$.
This sufficiency will be proved subsequently, first we discuss the probability of selecting such a code.
Codes which satisfy Equations~\eqref{eq:thm:-t:suff1} and~\eqref{eq:thm:-t:suff2} for all $(m,\tilde m) \in \mcf{M} \times \mcf{M}$ are chosen from the code construction process with probability greater than
\begin{equation}\label{eq:thm:-t:p(code)}
1- 4e^{-\frac{n^2}{4}+ 2nr}.
\end{equation}
Indeed, with probability $2^{-n \beta}$ a code is chosen such that $m \in \mcf{M}(k_2)$. 
Similarly with probability  
\[
2^{-n\beta} \frac{2^{n(r-\beta)}-1}{2^{nr} - 1} %= 2^{-2n\beta} \frac{2^{nr} - 2^{n\beta}}{2^{nr}-1} 
= 2^{-2n \beta} \frac{1 - 2^{n(\beta - r) }}{1 - 2^{-nr}},
\]
a code is chosen that satisfies $(m,\tilde m) \in \mcf{M}(k_2) \times \mcf{M}(k_2)$.
There are $2^{2n\beta}$ sets $\mcf{M}(k_2)$ that are independently chosen, and by requirement $\beta > 2 n^{-1}\logt n$.
Thus for a fixed $(m, \tilde m) \in \mcf{M}\times \mcf{M}$, with probability greater than $1-2e^{-n^2/4}$ a code is chosen 
\[
e^{-2} 2^{n \beta } < \abs{\set{k_2 :m \in \mcf{M}(k_2)}} \leq e^2 2^{n\beta }
\]
and with probability greater than $1-2e^{-n^2/4}$ a code is chosen such that 
\[
\abs{\set{k_2 :(m,\tilde m) \in \mcf{M}(k_2) \times \mcf{M}(k_2)} } \leq (ne)^2
\]
by Corollary~\ref{cor:ck}.
Equation~\eqref{eq:thm:-t:p(code)} is obtained by using the union bound to consider all $(m,\tilde m) \in \mcf{M} \times \mcf{M}$ simultaneously. 
%We now move to showing that these sufficient conditions guarantee Equation~\eqref{eq:thm:-t:tt1e}. 

Next, we show that codes that meet the sufficiency conditions imply the average type I error requirement, but first observe that
\begin{align}
&\sum_{k_2:  m \in \mcf{M}(k_2)} \frac{1}{\abs{\set{k_2 :  m \in \mcf{M}(k_2)}}} \omega_{f,\varphi'}(\psi,z^n,g_{k_2}^{-1}( m),k_1,k_2) \notag\\
&= \sum_{y^n} \psi(y^n|z^n) \sum_{ \tilde m \in \mcf{M} - \set{m} } \varphi( \tilde m  | y^n,k_1) \sum_{k_2:  m \in \mcf{M}(k_2)} \frac{1}{\abs{\set{k_2 :  m \in \mcf{M}(k_2)}}}  \idc{\tilde m \in \mcf{M}(k_2)}\label{eq:thm:-t:idca-2}\\
&\leq \sum_{y^n} \psi(y^n|z^n) \sum_{ \tilde m \in \mcf{M} - \set{m} } \varphi( \tilde m  | y^n,k_1) n^2 e^4 2^{-n \beta }\label{eq:thm:-t:idca-1}\\
&= \omega_{f,\varphi}(\psi,z^n,m,k_1)  n^2 e^4 2^{-n \beta},\label{eq:thm:-t:idca}
\end{align}
where~\eqref{eq:thm:-t:idca-2} is because $\varphi'(g_{k_2}^{-1}(m) | y^n,k_1) = \varphi(m|y^n,k_1) \idc{m \in \mcf{M}(k_2)}$, and~\eqref{eq:thm:-t:idca-1} is by Conditions~\eqref{eq:thm:-t:suff1} and~\eqref{eq:thm:-t:suff2}. 
The type I error for $(f',\varphi')$ is
\begin{equation}\label{eq:cc2:t1e:2}
\max_{\psi}\sum_{z^n,m,k_1,x^n} q^{\otimes n}(z^n|x^n) f(x^n| m,k_1) 2^{-n( \kappa + r + \beta )} \sum_{k_2 :  m \in \mcf{M}(k_2)}  \omega_{f'\varphi'}(\psi,z^n,g_{k_2}^{-1}(m),k_1,k_2) ,
\end{equation}
where it should be noted the summation is over $m \in \mcf{M}$ instead of $m' \in \mcf{M}'$.
But, if a code satisfies the sufficient conditions~\eqref{eq:thm:-t:suff1} and~\eqref{eq:thm:-t:suff2}, then
\begin{equation}\label{eq:cc2:t1e:3}
\sum_{k_2:  m \in \mcf{M}(k_2)}  \omega_{f',\varphi'}(\psi,z^n,g_{k_2}^{-1}( m),k_1,k_2) < n^2 e^6  \omega_{f,\varphi}(\psi,z^n,m,k_1) 
\end{equation}
by Equation~\ref{eq:thm:-t:idca}.
Thus~\eqref{eq:cc2:t1e:2} is less than 
\begin{equation}
n^2 e^6 2^{-n \beta}  \max_{\psi} \sum_{z^n,m,k_1,x^n} q^{\otimes n}(z^n|x^n) f(x^n| m,k_1) 2^{-n( \kappa + r)} \omega_{f,\varphi}(\psi,z^n,m,k_1) = n^2e^6 2^{-n \beta} \omega_{f,\varphi},
\end{equation}
whence Equation~\eqref{eq:thm:-t:tt1e}.

\subsection{Proofs of Lemma~\ref{lem:unique}, Lemma~\ref{lem:obvious}, Corollary~\ref{cor:ck}, and Lemma~\ref{lem:tbcut}}
\subsubsection{Proof of Lemma~\ref{lem:unique} }\label{app:unique}~\\
\begin{IEEEproof}
First note that 
\[
\nu(c|b) = p_{z^n|u^n}(c|b) = \sum_{a} p_{z^n|u^n,w^n}(c|b,a) p_{w^n|u^n}(a|b) = p_{z^n|u^n,w^n}(c|b,a_b)
\]
where $a_b$ is the value such that $p_{u^n|w^n}(b|a_b)= \sigma(b|a_b)$ is non-zero. Hence
\begin{align*}
p_{z^n|w^n}(c|a) &= \sum_{b}p_{z^n|u^n,w^n}(c|b,a) p_{u^n|w^n}(b|a) \\
&= \sum_{b: a_b = b }p_{z^n|u^n,w^n}(c|b,a) \sigma (b|a) + \sum_{b: a_b \neq b }p_{z^n|u^n,w^n}(c|b,a) \sigma(b|a) \\
&= \sum_{b: a_b = b } \nu(c|b)\sigma(b|a) \\
&= \nu\sigma(c|a).
\end{align*}
\end{IEEEproof}

\subsubsection{Proof of Lemma~\ref{lem:obvious}}\label{app:obvious} ~\\
\begin{IEEEproof}
If $U^n$ is chosen uniformly over $\mcf{T}^n_{\sigma}(w^n)$ then 
\begin{equation}\label{eq:lem:obvious:1}
\Pr \left( y^n \in \mcf{T}^n_{\mu}(U^n) \right) = \sum_{u^n} p_{U^n}(u^n) \idc{y^n \in \mcf{T}^n_{\mu}(u^n) } = \frac{\abs{\set{u^n : y^n \in \mcf{T}^n_{\mu}(u^n),~u^n \in \mcf{T}^n_{\sigma}(w^n)}} }{\abs{\mcf{T}^n_{\sigma}(w^n)}}.
\end{equation}
for any fixed $y^n$.
The set in the numerator of the RHS of Equation~\eqref{eq:lem:obvious:1} is equal to $\mcf{T}^n_{\bar \mu}(y^n,w^n)$ where $\bar \mu(b|a,c) \mu\sigma(c|a) = \mu(c|b) \sigma(b|a)$. Indeed, all $u^n$ in the numerator of Equation~\eqref{eq:lem:obvious:1} have a fixed empirical distribution $(y^n,u^n,w^n)$ since $u^n$ determines $w^n$ uniquely. Now consider expanding $p_{y^n,u^n|w^n}$ in two ways. First, 
\[
p_{y^n,u^n|w^n} = p_{y^n|u^n,w^n} \circ p_{u^n|w^n} = \mu \circ \sigma,
\]
since $u^n$ determines $u^n$.
Second,
\[
p_{y^n,u^n|w^n} = p_{u^n|y^n,w^n} \circ p_{y^n|w^n} = \bar\mu \circ \mu\sigma,
\]
because $p_{u^n|y^n,w^n} = \bar \mu$ by definition, and $p_{y^n|w^n} = \mu \sigma$ by Lemma~\ref{lem:unique}. Equating the two expansions and solving for $\bar \mu$ gives $\bar \mu(c|b,a) = \frac{\mu(c|b)\sigma(b|a)}{\mu\sigma(c|a)}$.

Thus 
\begin{equation}
\Pr \left( y^n \in \mcf{T}^n_{\mu}(U^n) \right) = 2^{n \left( \msf{H}(\bar \mu | \mu\sigma \circ \tau) - \msf{H}(\sigma|\tau) \right) + O(\logt n)} = 2^{-n\msf{I}(\mu,\sigma|\tau) + O(\logt n)}
\end{equation}
follows from substituting $\mcf{T}^n_{\bar \mu}(y^n,w^n)$ into the numerator of Equation~\eqref{eq:lem:obvious:1} and because
\begin{align*}
&\msf{H}(\bar \mu | \mu\sigma \circ \tau) - \msf{H}(\sigma|\tau) \\
&= - \sum_{c,b,a} \bar \mu(b|a,c) \mu\sigma(c|a) \tau(a) \logt \bar \mu(b|a,c)  +  \sum_{c,b,a}  \mu(c|b) \sigma(b|a) \tau(a) \logt \sigma(b|a)\\
%&= \sum_{c,b,a}  \mu(c|b) \sigma(b|a) \tau(a) \logt \frac{\sigma(b|a)}{\bar \mu(b|a,c) } \\
&= -\sum_{c,b,a}  \mu(c|b) \sigma(b|a) \tau(a) \logt \frac{\mu(c|b)}{ \mu\sigma(c|a) } \\
&= -\msf{I}(\mu,\sigma|\tau).
\end{align*}

\end{IEEEproof}

\subsubsection{Proof of Corollary~\ref{cor:ck}}\label{app:ck}~\\
\begin{IEEEproof}
Let $V_1, \dots, V_{2^{n \beta + f(n)}}$ be IID Bern($2^{-n \rho + g(n)}$) RVs. Define $\Delta(n) = \abs{f(n) - g(n)}$. If $n\beta - n\rho + f(n) - g(n) \geq 2 \logt n$ then
\begin{align}
\Pr \left( \sum_{i} V_i < \left \lfloor e^{-1} 2^{n \left| \beta -\rho \right|^+ - \Delta(n) } \right \rfloor \right) &\leq \Pr \left( \sum_{i} V_i < e^{-1} 2^{n( \beta -\rho) + f(n) - g(n) } \right) \notag \\
&< e^{-(1 - 2e^{-1})2^{n (\beta - \rho) + f(n) - g(n)}} < e^{-n^2/4}\label{cor:2}
\end{align}
and
\begin{equation}\label{cor:1}
\Pr \left( \sum_{i} V_i > e 2^{n \left| \beta -\rho \right|^+ + \Delta(n) } \right) \leq \Pr \left( \sum_{i} V_i > e 2^{n \beta -\rho + f(n) - g(n) } \right) < e^{-2^{n (\beta - \rho) + f(n) - g(n)}} < e^{-n^2}
\end{equation}
by Lemma~\ref{lem:ck}.

On the other hand, if $n\beta - n\rho + f(n) - g(n) < 2 \logt n$ then
\begin{align}
\Pr \left( \sum_{i} V_i < \left \lfloor n^{-2} 2^{n \left| \beta -\rho \right|^+ - \Delta(n) } \right \rfloor \right) &\leq \Pr \left( \sum_{i} V_i < 0 \right) 
%\notag \\ &
= 0 \label{cor:3}
\end{align}
and
\begin{align}
\Pr \left( \sum_{i} V_i > n^2 e^2 2^{n \left| \beta -\rho \right|^+ + \Delta(n) } \right) &\leq \Pr \left( \sum_{i} V_i < n^2 e^{2}  \right) \notag \\
&= \Pr \left( \sum_{i} V_i < \left( \frac{n^2 e^{2}}{2^{n(\beta- \rho) + f(n) - g(n)}} \right)   2^{n(\beta- \rho) + f(n) - g(n)} \right)   \notag \\
&< e^{-n^2e^2} \label{cor:4}
\end{align}
by Lemma~\ref{lem:ck}, and in particular because
\begin{align*}
c\left( \frac{n^2 e^2}{2^{n(\beta- \rho) + f(n) - g(n)}} \right) 2^{n(\beta- \rho) + f(n) - g(n)} &= n^2e^2 \ln \frac{n^2 e^2}{2^{n(\beta- \rho) + f(n) - g(n)}}   - n^2e^2 + 2^{n(\beta- \rho) + f(n) - g(n)}\\
&> n^2e^2 \ln \frac{n^2 e^2}{n^2}   - n^2e^2 \\
&= n^2 e^2.
\end{align*}

\end{IEEEproof}

\subsubsection{Proof of Lemma~\ref{lem:tbcut}}\label{app:tbcut}~\\

\begin{IEEEproof}

Write $\abs{\cup_{i} \mcf{A}_i}$ as the summation of its elements, that is
\begin{equation}
\abs{\cup_{i} \mcf{A}_i} = \sum_{x } \idc{x\in \cup_{i} \mcf{A}_i}.
\end{equation}
From this equation it must also follow that
\begin{equation}
\abs{\cup_{i} \mcf{A}_i} = \sum_{x} \sum_{i} \frac{\idc{x \in \mcf{A}_i}}{\abs{\set{i : x \in \mcf{A}_i}}}
\end{equation}
since
\[
\idc{x\in \cup_{i} \mcf{A}_i} = \sum_i \frac{\idc{x \in \mcf{A}_i}}{\abs{\set{i : x \in \mcf{A}_i}}}
\]
for every $x \in \cup_i \mcf{A}_i$. Swapping the order of the summations and replacing the denominator with the min or max results in the Lemma. 
\end{IEEEproof}

}{}

\end{document}